\documentclass[aps,prb,twocolumn,superscriptaddress]{revtex4-2}
\usepackage[T1]{fontenc}
\usepackage[latin9]{inputenc}
\usepackage{babel}
\usepackage{amsmath}
\usepackage{amssymb}
\usepackage{stmaryrd}
\usepackage{graphicx}
\usepackage{wasysym}
\usepackage[unicode=true]{hyperref}
\usepackage{breakurl}



\begin{document}

\title{Wave-packet dynamics in pseudo-Hermitian lattices: Coexistence of
Hermitian and non-Hermitian wavefronts}

\author{Alon Beck\footnote{alonbk2@gmail.com}
and Moshe Goldstein}

\affiliation{Raymond and Beverly Sackler School of Physics and Astronomy, Tel
Aviv University, Tel Aviv 6997801, Israel}
\date{\today}
\begin{abstract}
This paper investigates wave-packet dynamics in non-Hermitian lattice systems and reveals a surprising phenomenon: The simultaneous propagation of two distinct wavefronts, one traveling at the non-Hermitian velocity and the other at the Hermitian velocity. We show that this dual-front behavior arises naturally in systems governed by a pseudo-Hermitian Hamiltonian. Using the paradigmatic Hatano-Nelson model as our primary example, we demonstrate that this coexistence is essential for understanding a wide array of unconventional dynamical effects, including abrupt ``non-Hermitian reflections'', sudden shifts of Gaussian wave-packets, and disorder-induced emergent packets seeded by the small initial tails. We present analytic predictions that closely match numerical simulations. These results may offer new insight into the topology of non-Hermitian systems and point toward measurable experimental consequences.
\end{abstract}
\maketitle

\section{Introduction}

Non-Hermitian systems \citep{Bender1998,Bender2007,Rotter2009,ElGanainy2018,Ashida2020}
have attracted significant interest in recent years, spanning a wide
range of applications \textemdash{} from classical mechanical \citep{Brandenbourger2019,Ghatak2020,Wang2023},
acoustic \citep{Fleury2015,Ding2016,Zhu2018,Wang2018,Zhang2021,Gu2021,Gao2021,Zhang2023},
and photonic waves \citep{Ding2015,Longhi2015,Zhen2015,Feng2017,Pan2018,Oezdemir2019}
with gain and loss to quantum atomic \citep{Peng2016,Li2019,Takasu2020,Liang2022}
and electronic systems \citep{Schindler2011,DeCarlo2022,Yuan2022}.
They can challenge the conventional understanding of fundamental
concepts. For example, the non-Hermitian skin effect \citep{Shen2018,MartinezAlvarez2018,Lee2019,Li2020,Okuma2020,Borgnia2020,Zhang2020,Zhang2021a,Zou2021,Zhang2022,Lin2023},
which is characterized by the exponential localization of bulk states
at the system's boundaries, disrupts the traditional bulk-edge correspondence
\citep{Kunst2018,Shen2018a,Xiao2020,Rapoport2023}. In addition, the
conventional Bloch-band theory must be extended with a generalized
Brillouin zone to adequately represent their topological behaviors
\citep{Yao2018,Yokomizo2019,Bergholtz2021}.

While most studies focus on eigenvalue properties, it has recently
become clear that the dynamics of these systems can also play a significant
role in understanding their properties \citep{Li2022,Longhi2022,Li2024,Li2022a}.
Related semiclassical and quantum-geometric aspects of non-Hermitian wave-packet dynamics have also been explored in Refs.~\citep{Silberstein2020,BarAlon2024,baralon2025,behrends2025}.
Given the complexity of the spectrum, the system's time evolution
can exhibit non-trivial behaviors, such as the dominance of specific
eigenstates due to non-Hermitian skin effect \citep{Longhi2019},
and peculiar wave-packet behavior near the boundaries \citep{Li2022}.

In this paper, we will investigate in particular the dynamics of systems
described by a pseudo-Hermitian Hamiltonian \citep{Lee1969,MOSTAFAZADEH2010}.
Such Hamiltonians have the property $H^{\dagger}=\eta^{-1}H\eta,$
where $\eta$ is a positive-definite operator. This class includes,
in particular, all diagonalizable Hamiltonians with parity-time (PT)
symmetry \citep{Mostafazadeh2002,Mostafazadeh2002a,Mostafazadeh2002b,Zhang2020a}.
We will show that such systems presents not only the typical exponentially
increasing wave-packet dynamics \citep{Longhi2019}, but also a \textit{Hermitian}
wave-packet, co-existing with the former. 
This coexistence arises naturally from the pseudo-Hermiticity of the Hamiltonian and leads to Hermitian wave-packets that reflect from the boundaries-behavior that, in turn, governs the boundary dynamics of the non-Hermitian wave-packet.
To demonstrate the properties discussed above, we will
focus primarily on the simple Hatano-Nelson model, analyzing it both
numerically and analytically. However, these results can also be generalized
to other pseudo-Hermitian systems.

The rest of the paper is organized as follows: In Section \ref{sec:Deriving-the-dynamics},
we begin by demonstrating how the existence of a transformation that
makes the Hamiltonian Hermitian (a condition equivalent to pseudo-Hermiticity
\citep{Jones2005}) implies the coexistence of two types of dynamics:
Hermitian and non-Hermitian. We show how this transformation is derived
within the Hatano-Nelson model.

In Section \ref{sec:Non-Hermitian-wave-packet-dynami},
we investigate the non-Hermitian dynamics of Gaussian wave-packets,
focusing on cases with small ($\sigma\ll a$) and moderate ($a<\sigma\ll L$)
wave-packet width $\sigma$ ($a$ and $L$ being the lattice spacing and system size), as detailed in Sections \ref{sec:single_site_starting_condition_dynamics}
and \ref{subsec:Gaussian-wave-packets-()}, respectively. While the
coexistence of the two wave-packets is observed in both scenarios,
the case of $a<\sigma\ll L$ can exhibit additional, unexpected behavior, such as:
(a) an abrupt change in velocity resulting from the reflection of the Hermitian wave-packet at the wall; and
(b) a sudden spatial ``jump'' of the wave-packet, caused jointly by this reflection and by the finite maximal velocity on the lattice. The latter effect produces two wavefronts that propagate side by side --- at a certain point, the reflected front can overtake the original one, giving rise to the striking apparent "jump" in the wave-packet's position.

In Sec.~\ref{subsec:Disorder}, we examine the effect of weak disorder, showing that it can likewise induce a transition --- an abrupt change in the wave-packet's position --- but now instead of arising from reflection, it originates from the small initial tails generated by the disorder.

Finally, in Sec. \ref{sec:Beyond-the-Hatano-Nelson},
we discuss how our results can be applied to the non-Hermitian SSH
model \citep{Su1979,Lieu2018,Kunst2018}, demonstrating that our findings extend to other pseudo-Hermitian Hamiltonians. We conclude by
examining the implications of our results and outlining directions
for future research. Some technical details are relegated to the Appendixes.

\section{\label{sec:Deriving-the-dynamics}Deriving dynamics via local transformation}

\subsection{Dynamics via transformation}
Consider a pseudo-Hermitian
Hamiltonain $H$, where $H^{\dagger}=\eta^{-1}H\eta$ for some positive-definite
$\eta$. By choosing $S=\sqrt{\eta}$, we can transform the Hamiltonian
to its Hermitian counterpart
\begin{equation}
H'=S^{-1}HS,\label{eq:transformation_generalized}
\end{equation}
where $H'=H'^{\dagger}$. We can notice now that the time-development
of any wavefunction $\psi$ can be written as \citep{Li2022}
\begin{multline}
\left\langle m|e^{-iHt}|\psi(0)\right\rangle =\left\langle S^{\dagger}m\left|e^{-iH't}\right|S^{-1}\psi(0)\right\rangle ,\label{eq:dynamics_using_transformation_generalized}
\end{multline}
where $H'=S^{-1}HS$. This result allows us to obtain the dynamics by simply calculating
the time-development of $S^{-1}\psi(0)$ in respect to the \textit{Hermitian}
Hamiltonian $H'$, and using $S^{\dagger}$ again at the end. Since
it is easier to calculate the propagator of a Hermitian system, Eq.
(\ref{eq:dynamics_using_transformation_generalized}) offers a valuable
method to calculate the dynamics of the non-Hermitian system. However,
our main observation is more subtle: Eq. (\ref{eq:dynamics_using_transformation_generalized})
actually reveals the \textit{coexistence} of two distinct types of
dynamics \textemdash{} Hermitian and non-Hermitian. We will now investigate
this coexistence in details within the Hatano-Nelson model.

\subsection{The model}The paradigmatic Hatano-Nelson Hamiltonian \citep{Hatano1996}
is the simplest model which displays most of the peculiarities of
the non-Hermitian dynamics studied here. It is a 1D tight binding
model with non-reciprocal hopping, given by
\begin{equation}
H=\sum_{n}t_{l}\left|n\right\rangle \left\langle n+1\right|+t_{r}\left|n+1\right\rangle \left\langle n\right|,\label{eq:Hanato_nelson}
\end{equation}
where $n=1,...,N$. We assume that $t_{l}$ and $t_{r}$ are real and share the same sign,
so that the Hamiltonian is pseudo-Hermitian \footnote{Adding phases of \( e^{i\varphi} \) and \( e^{-i\varphi} \) to \( t_r \) and \( t_l \), respectively, preserves pseudo-Hermiticity, while any global phase does not.}. For periodic boundary conditions (PBC), the wavefunctions are the
well known Bloch waves $\left\langle \left.n\right|k_{m}\right\rangle =\frac{1}{\sqrt{N}}e^{ikan}$,
where $a$ is the lattice constant and $k_{m}=\frac{2\pi}{a}\frac{m}{N}$,
$m=0,1,...,N-1$. The energies are given by
\begin{equation}
E(k)=\left(t_{l}+t_{r}\right)\cos(ka)+i\left(t_{l}-t_{r}\right)\sin(ka).\label{eq:Hatano-Nelson-specturm}
\end{equation}
For open boundary conditions (OBC), the eigenstates and spectrum are
given by
\begin{equation}
\left\langle \left.n\right|\psi^{(m)}\right\rangle =r^{n}\left(e^{i\theta_{m}n}-e^{-i\theta_{m}n}\right),\,E_{m}=2\sqrt{t_{r}t_{l}}\cos\left(\theta_{m}\right),\label{eq:OBC_eigenfunctions}
\end{equation}
where $r=\sqrt{\frac{t_{r}}{t_{l}}}$, $\theta_{m}=\pi\frac{m}{N+1}$
\citep{Yokomizo2019} (where we assumed $t_{l}>t_{r}>0$). We note
that for OBC all of the eigenstates are localized on one the left
edge of the system. This feature is the well-known non-Hermitian skin
effect \citep{Shen2018,MartinezAlvarez2018,Lee2019,Li2020,Okuma2020,Borgnia2020,Zhang2020,Zhang2021a,Zou2021,Zhang2022,Lin2023},
arising due to asymmetric couplings.

As was shown above, the spectrum and the eigenstates in the cases
of PBC and OBC are very different. However, it should be noted that
if the system is large enough, the locality of the Hamiltonian implies
that the dynamics of the bulk of the system is independent of the
boundary conditions, which affect only wave-packets approaching the boundaries.
We will focus on the case of OBC, which we show below to be
pseudo-Hermitian, and whose boundary structure gives rise to intriguing
phenomena in the dynamics in their vicinity, as discussed later.
In this case, the localization of the wavefunction in Eq. (\ref{eq:OBC_eigenfunctions})
can be ``undone'' using a diagonal non-unitary similarity transformation
\citep{Yao2018}
\begin{equation}
\left\langle n\left|S\right|m\right\rangle =\delta_{n,m}r^{n},\label{eq:Transformation}
\end{equation}
where $\delta_{n,m}$ is the Kronecker delta and $r=\sqrt{\frac{t_{r}}{t_{l}}}$
as before. After the transformation (\ref{eq:transformation_generalized}),
the Hamiltonian becomes
\begin{equation}
H'=\sqrt{t_{l}t_{r}}\sum_{n}\left|n\right\rangle \left\langle n+1\right|+\left|n+1\right\rangle \left\langle n\right|,\label{eq:Hamiltonain_Transformed}
\end{equation}
that is, we get a simple tight-binding Hermitian Hamiltonian with
a hopping amplitude $\sqrt{t_{l}t_{r}}$. We note that this also implies
that the Hamiltonian is pseudo-Hermitian by choosing $\eta=SS^{\dagger}$.

Using a transformation $S$ of the type of Eq. (\ref{eq:Transformation}),
and using the observation from Eq. (\ref{eq:dynamics_using_transformation_generalized}),
the time-development can be written as

\begin{multline}
\left\langle m|e^{-iHt}|\psi(0)\right\rangle =r^{m}\left\langle m\left|e^{-iH't}\right|S^{-1}\psi(0)\right\rangle \label{eq:dynamics_using_transformation}
\end{multline}
That is, the non-Hermitian wave-packet can be obtained by calculating
first the Hermitian dynamics of $S^{-1}\psi(0)$ and multiplying by
the exponential factor $r^{m}$ in the end. Therefore, the the non-Hermitian
wave-packet is a manifestation of the \textit{tail} of the Hermitian
wave-packet. As we will see, the velocity and exponential growth of
the non-Hermitian wave-packet peak make it more prominent than its Hermitian counterpart.
However, if we look at some fixed point at space, the behavior
in late times would usually be governed by the Hermitian wave-packet
dynamics. In addition, only the Hermitian wave-packet is capable of
reflecting from the edges of the system (or potential barriers); this
fact would prove to be crucial for the analysis of wave-packet dynamics near
the edges.

\section{\label{sec:Non-Hermitian-wave-packet-dynami}Non-Hermitian wave-packet
dynamics}

We will now examine and demonstrate the differences between the two
coexisting dynamics in the Hatano-Nelson model with OBC. The initial
conditions of the wave-packet will be taken as a Gaussian with width $\sigma$ and average quasi-momentum $k_{0}$, focusing on two cases:
$\sigma\ll a$ (equivalent to a delta-function $\left\langle \left.n\right|\psi(t=0\right\rangle =\delta_{n,n_{0}}$)
and $a<\sigma\ll L$, where $L=Na$. We will demonstrate the presence of coexisting
dynamics in both cases and show that certain intriguing phenomena
(such as emerging wave-packets) are observed only in the second case,
when the width exceeds a critical threshold $\sigma>\sigma_{c}$.
Finally, we will show that these results extend beyond the HN model,
provided the Hamiltonian is pseudo-Hermitian.

\subsection{\label{sec:single_site_starting_condition_dynamics}Single-site initial
condition ($\sigma\ll a$)}

Here we will focus on the case where the initial condition is $\left|\psi(t=0)\right\rangle =\left|n\right\rangle $,
with $n$ representing a lattice site. We will demonstrate the coexistence
of two wave-packets, showing that only the Hermitian one is responsible
for the oscillations that are being observed in the system. We begin
with an analytical investigation of the dynamics. Using Eq. (\ref{eq:dynamics_using_transformation_generalized}),
we can write
\begin{equation}
\left\langle m\right.\left|\psi(t)\right\rangle =r^{m-n}\left\langle m\left|e^{-iH't}\right|n\right\rangle ,\label{eq:dynamics_of_delta_starting_condition}
\end{equation}
where the Hermitian Hamiltonain $H'$ is given by Eq. (\ref{eq:Hamiltonain_Transformed}).
The propagator of the Hemitian Hamiltonian is given by
\begin{equation}
\left\langle m\left|e^{-iH't}\right|n\right\rangle =\frac{1}{N}\sum_{k}e^{ika(m-n)}e^{-2it_{0}\cos(ka)t},\label{eq:hermitian_propagator}
\end{equation}
where $t_{0}=\sqrt{t_{r}t_{l}}$ is the hopping amplitude of $H'$.
If $N$ is large, we can replace the sum by an integral and get
\begin{equation}
\left\langle m\right.\left|\psi(t)\right\rangle =\frac{1}{2\pi}r^{m-n}\left(-i\right)^{m-n}J_{m-n}(2t_{0}t),\label{eq:Bessel_function_propagator}
\end{equation}
where $J_{n}(x)=\frac{1}{2\pi}\int_{-\pi}^{\pi}e^{i(ny-x\sin y)}dy$
is the Bessel function of the first kind. We can now calculate the
velocities of the Hermitian and non-Hermitian wave-packets: In the
case of the Hermitian Hamiltonian $H'$, the maximal velocity $v_{h}$
is given as usual by $\max\left(\frac{\partial E}{\partial k}\right)$,
which is obtained for $k_{m}=\pm\frac{\pi}{2a}$. Therefore,
\begin{equation}
v_{h}=\pm2a\sqrt{t_{r}t_{l}}.\label{eq:Hermitian_velocity}
\end{equation}
However, since the non-Hermitian spectrum (given by Eq. \ref{eq:Hatano-Nelson-specturm})
is complex, the amplitude of each eigenstate obtains a factor of $e^{\mathrm{Im}(E)t}$,
making only $\max\left(\mathrm{Im}E\right)$ significant at long times.
Therefore, the non-Hermitian velocity is given by $v_{\mathrm{nh}}=\mathrm{Re} \left. \frac{\partial E}{\partial k} \right|_{k = k_m},$
for $k_{m}$ that maximize $\mathrm{Im}E(k)$ \citep{Longhi2019}.
We can see that in this case $k_{m}=-\pi/2$ also, and therefore
\begin{equation}
v_{\mathrm{nh}}=-a(t_{r}+t_{l}).\label{eq:nonHermitian_velocity}
\end{equation}
We note that for any values of $t_{r},t_{l}$, the non-Hermitian velocity
will be larger than the Hermitian one.

We now proceed to numerical simulations, obtained by computation of
the propogator $e^{-iHt}$. We note that the computer precision may
introduce an effective disorder term. Due to the non-Hermitian nature
of the system, even weak disorder can influence the dynamics
at late times (for more details see Sec. \ref{subsec:Disorder}).
To avoid such terms, one needs to increases the precision \textemdash{}
in Matlab, for example, this can be achieved using the arbitrary precision
library Advanpix \citep{Advanpix}.
\begin{figure}
\centerline{\includegraphics[scale=0.6]{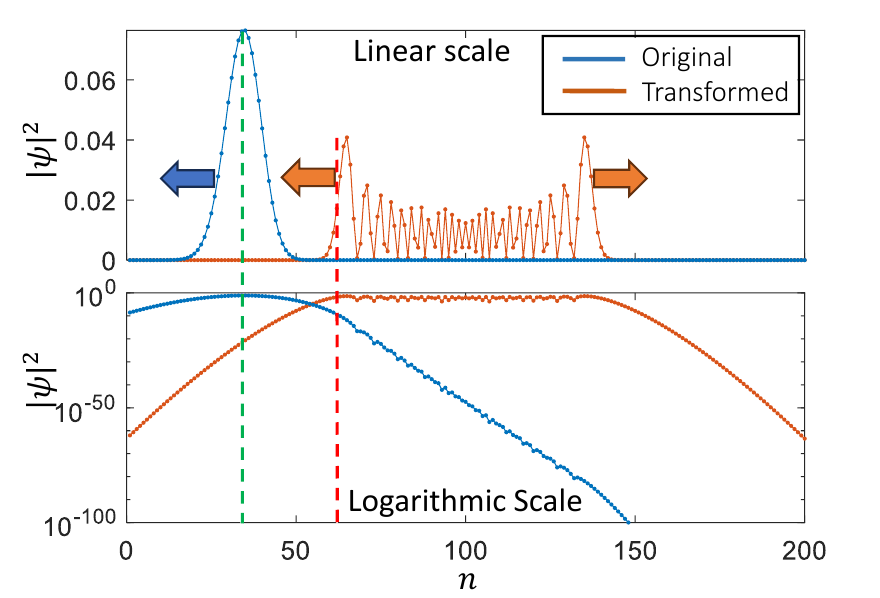}}\caption{\label{fig:delta_dynamics}Snapshot of the non-Hermitian dynamics
with and without the transformation (the arrows indicate the direction
of propagation of the wave-packets). See Video\_1 in the Supplementary Material for an animation of the dynamics.
Blue dataset: A wave-packet with initial conditions $\left\langle n\right.\left|\psi(t=0)\right\rangle =\delta_{n,x_{0}}$,
where $x_{0}=N/2$, evolved numerically to time $t=60$ with the non-Hermitian
Hamiltonian, with $t_{l}=2,t_{r}=0.2$ (OBC). For convenience, the
wave-packet is normalized to 1 (otherwise, it would present an exponential
growth). Orange dataset: the same data after the transformation $y_{n}\rightarrow y_{n}r^{-n}$
(and normalization). The green/red dotted lines are the predicted
positions of the fronts of the Hermitian and non-Hermitian wave-packets:
$x_{0}-v_{\mathrm{nh}}t$ (green) and $x_{0}-v_{h}t$ (red). The number
of sites is $N=200$, and the lattice spacing is $a=1$.}
\end{figure}
Fig. \ref{fig:delta_dynamics} presents an example of a wave-packet
with the initial condition $\left\langle n\right.\left|\psi(t=0)\right\rangle =\delta_{n,x_{0}}$,
where $x_{0}=N/2=100,$ propagated numerically (up to the time given in the caption),
with and without the transformation, Eq. (\ref{eq:Transformation}).
We can see that without the transformation (blue data), the non-Hermitian
system exhibits a wave-packet moving to the left (with exponentially-increasing
amplitude, which is not apparent here only since the wave-packet is
normalized). With the transformation (orange data), it is clear that
the dynamics becomes Hermitian, displaying a symmetrical behavior
of two wave-packets moving in opposite directions (the oscillations
in between are the typical behavior we get from the Bessel-function propagator
\citep{Schoenhammer2019}). The green/red dotted lines represents
the predicted position of the wave-packets due to non-Hermitian/Hermitian
velocities, given by Eq. (\ref{eq:nonHermitian_velocity}), (\ref{eq:Hermitian_velocity}),
respectively. We can gain more insights by looking at the same picture
in log-scale (bottom panel); we see that the Hermitian wave-packet
does not disappear after the transformation, but can be still found
in the tail of the non-Hermitian wave-packet.

\begin{figure}
\centerline{\includegraphics[scale=0.5]{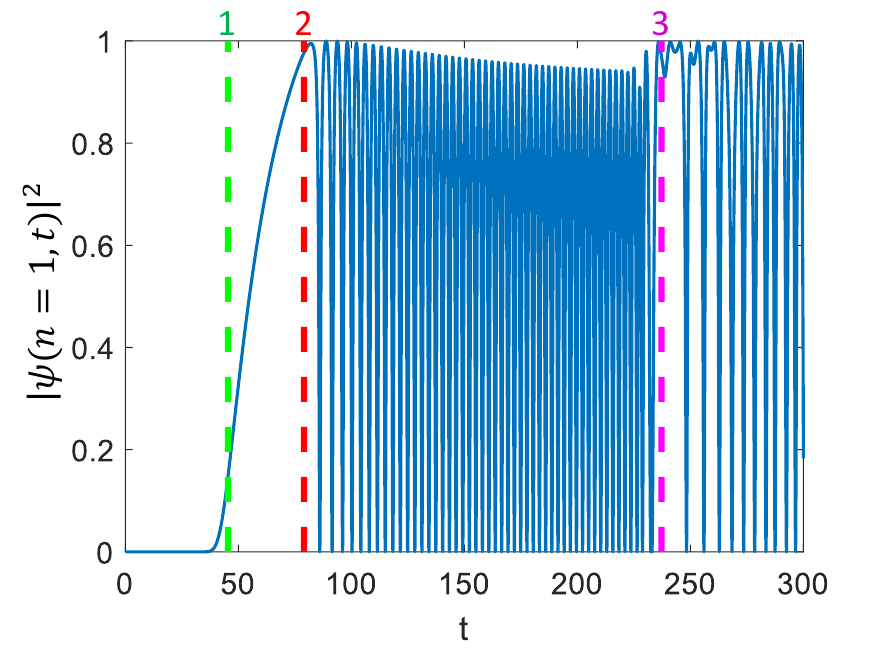}}\caption{\label{fig:delta_dynamics_of_first_site}Numerical results of the
amplitude at the leftmost site as a function of time in the same setup
as Fig. \ref{fig:delta_dynamics}. Recalling that $x_{0}=N/2$, three
timestamps are marked by $t_{1}=\frac{x_{0}}{v_{\mathrm{nh}}}\approx45$
(green), indicating the time when the non-Hermitian wave-packet hits
the edge, $t_{2}=\frac{x_{0}}{v_{\mathrm{h}}}\approx79$ (red),
when the Hermitian wave-packet hits the edge, and $t_{3}=\frac{2N-x_{0}}{v_{\mathrm{nh}}}\approx237$
(purple), when the Hermitian wave-packet (that went to the other
side) hits the edge after being reflected from the other end.}
\end{figure}
In Fig. \ref{fig:delta_dynamics_of_first_site} we provide another evidence
for the significance of the Hermitian wave-packet in the non-Hermitian dynamics, by tracking the amplitude
of the wave-packet at the left edge. At time $t_{1}$, the non-Hermitian
wave-packet reaches the edge at first and sticks to it. Later, at
time $t_{2}$, the front of the Hermitian wave-packet also arrives
and becomes more dominant. Finally, at time $t_{3}$, we can also
see the front of the Hermitian wave-packet that went to the other
side and was reflected from the opposite edge. Using the non-Hermitian
and Hermitian velocities given by Eq. (\ref{eq:nonHermitian_velocity}),
(\ref{eq:Hermitian_velocity}), and recalling that the initial position
is the center ($N/2$), we can predict that $t_{1}=\frac{N/2}{v_{\mathrm{nh}}}$,
$t_{2}=\frac{N/2}{v_{\mathrm{h}}}$, and $t_{3}=\frac{3N/2}{v_{\mathrm{nh}}}$,
which fit well to the numerical results as shown in the figure. For
demonstration, see also the animation of Fig. \ref{fig:delta_dynamics}.

To sum up, in the above analysis we have shown that a simple initial
condition of a delta-function creates a non-Hermitian wave-packet
moving to the left (since $t_{l}>t_{r})$ with velocity $v_{\mathrm{nh}},$
and a Hermitian wave-packet with lower velocity $v_{h}$ living alongside
it. At early times, the non-Hermitian wave-packet is larger by several
orders of magnitude than the Hermitian wave-packet (making the latter
visible only in log-scale). However, if we probe the edge at later
times, the Hermitian wave-packet will eventually arrive and govern
the behavior at the boundary. Moreover, in contrast to the non-Hermitian
wave-packet that propagates only in a single direction, the Hermitian
wave-packets are capable of reflecting from the edges. We can understand
this by recalling that the propagator is obtained from Hermitian dynamics
multiply by an exponential factor, as shown by Eq. (\ref{eq:dynamics_of_delta_starting_condition}).
It is worth noting that even though any dynamics on the right edge
is suppressed by the exponential factor, it remains relevant since
the Hermitian wave-packets can probe it: While the wave-packet moving
to the right is indeed exponentially suppressed, when it reaches the
right boundary, it will switch direction and start being amplified
back as it moves to the left (until it reaches its maximum value on
the left boundary, see timestamp $t_{3}$ in Fig. \ref{fig:delta_dynamics_of_first_site}
for example).

\subsection{\label{subsec:Gaussian-wave-packets-()}Gaussian wave-packets ($a<\sigma\ll L$)}

We will now investigate the second case, where the initial conditions are Gaussian. First, we will derive a general expression for the position
of the wave-packet in the bulk of the lattice and compare it to the
result in the continuum. Next, we will demonstrate that an ``emergent
wave-packet'' can be observed due to reflection from the wall, and
show that the \textit{Hermitian} wave-packet is the one that determines
the time it occurs. Finally, we will discuss the critical threshold
$\sigma_{c,\mathrm{ref}}$, above which such phenomena can be observed.

\subsubsection{Dynamics in the bulk}

We consider a Gaussian wave-packet initial condition:
\begin{equation}
\left|\psi(t=0)\right\rangle =\sum_{n}\frac{1}{\sqrt{4\pi\sigma^{2}}}e^{-\frac{a^{2}\left(n-n_{0}\right)^{2}}{4\sigma^{2}}+ik_{0}an}\left|n\right\rangle .
\end{equation}
First, we will use Eq. (\ref{eq:dynamics_using_transformation_generalized})
to obtain an analytical expression for the dynamics. By completing
the square we get that
\begin{equation}
\left|S^{-1}\psi(0)\right\rangle =C\sum_{n}\frac{1}{\sqrt{4\pi\sigma^{2}}}e^{-\frac{a^{2}\left(n-\tilde{n}_{0}\right)^{2}}{4\sigma^{2}}+ik_{0}an}\left|n\right\rangle ,\label{eq:modified_psi_0_after_transformation}
\end{equation}
where 
\begin{equation}
C=\exp\left(\frac{\sigma^{2}}{a^{2}}\ln^{2}r-n_{0}\ln(r)\right),\,\tilde{n}_{0}=n_{0}-\frac{2\sigma^{2}}{a^{2}}\ln r.\label{eq:shifted_expectation_value}
\end{equation}
That is, after the transformation we obtain (up to some prefactor) a
Gaussian with the same width but a shifted expectation
value $\tilde{n}_{0}$. Therefore, to proceed we need to compute the
Hermitian dynamics of the Gaussian $\left|S^{-1}\psi(0)\right\rangle $
in respect to the Hamiltonian $H'$ in Eq. (\ref{eq:Hamiltonain_Transformed})
and then use the inverse transformation. In Appendix \ref{appendix_Hermitian-dynamics-of-gaussian}
we use the saddle point approximation (for large times, $t\gtrsim1/\sqrt{t_{r}t_{l}}$)
to derive an expression for the Hermitian dynamics of a Gaussian on
a lattice. Using Eq. (\ref{eq:eq:appendix_final_approximation_simpler}),
we find that
\begin{equation}
e^{-iH't}\left|S^{-1}\psi(0)\right\rangle \approx\tilde{C}\sum_{m}e^{-\frac{\sigma^{2}}{a^{2}}\left(\arcsin\left(\frac{a\left(m-\tilde{n}_{0}\right)}{tv_{h}}\right)-ak_{0}\right)^{2}}\left|m\right\rangle ,
\end{equation}
where $\tilde{C}$ is some constant and $v_{h}=2a\sqrt{t_{r}t_{l}}$
is the maximum Hermitian velocity as defined before. Finally, we get
\begin{equation}
\left\langle m|e^{-iHt}|\psi(0)\right\rangle \approx\tilde{C}r^{m}e^{-\frac{\sigma^{2}}{a^{2}}\left(\arcsin\left(\frac{a\left(m-\tilde{n}_{0}\right)}{tv_{h}}\right)-ak_{0}\right)^{2}}.\label{eq:non_hermitian_approximation}
\end{equation}
To analyze the approximation, we can expand it in powers of $t\frac{v_{h}}{a}$. In Appendix \ref{appendix_sbusec_small_and_large_t}
we find that

\begin{equation}
m_{\max}\approx\tilde{n}_{0}+2\sqrt{t_{r}t_{l}}\sin(ak_{0})t+\frac{2t_{r}t_{l}}{(\sigma/a)^{2}}\cos^{2}(ak_{0})\ln(r)t^{2},\label{eq:max_position_on_lattice}
\end{equation}
where $m_{\max}$ is the position of the maximum of the non-Hermitian
wave-packet. That is, to second order the wave-packet is just moving
with the initial $k_{0}$-dependent velocity $2\sqrt{t_{r}t_{l}}\sin(ak_{0})$,
while exhibiting an acceleration of $\frac{4t_{r}t_{l}}{(\sigma/a)^{2}}\cos^{2}(ak_{0})\ln(r)$. 

An analysis of the continuum limit ($a\rightarrow0$) is presented
in Appendix \ref{subsec:appendix_The-continuum-limit}; we show that
the non-Hermitian Hamiltonian reads 
\begin{equation}
H=E_{0}-\frac{1}{2m}\partial_{x}^{2}+b\partial_{x},
\end{equation}
where
\begin{equation}
m=\frac{1}{2a^{2}\sqrt{t_{l}t_{r}}},\,b=a\ln\left(\frac{t_{r}}{t_{l}}\right)\sqrt{t_{l}t_{r}},\label{eq:m_and_b}
\end{equation}
In addition, terms beyond $\left(t\frac{v_{h}}{a}\right)^{2}$ in
the expansion of Eq. (\ref{eq:non_hermitian_approximation}) vanish,
and Eq. (\ref{eq:max_position_on_lattice}) becomes
\begin{equation}
x_{\max}(t)=x_{0}+\frac{k_{0}}{m}t+\frac{b}{m\sigma^{2}}\frac{t^{2}}{2},\label{eq:max_position_continuum}
\end{equation}
in agreement with the result of Li and Wan \citep{Li2022}.

\begin{figure}
\centerline{\includegraphics[scale=0.5]{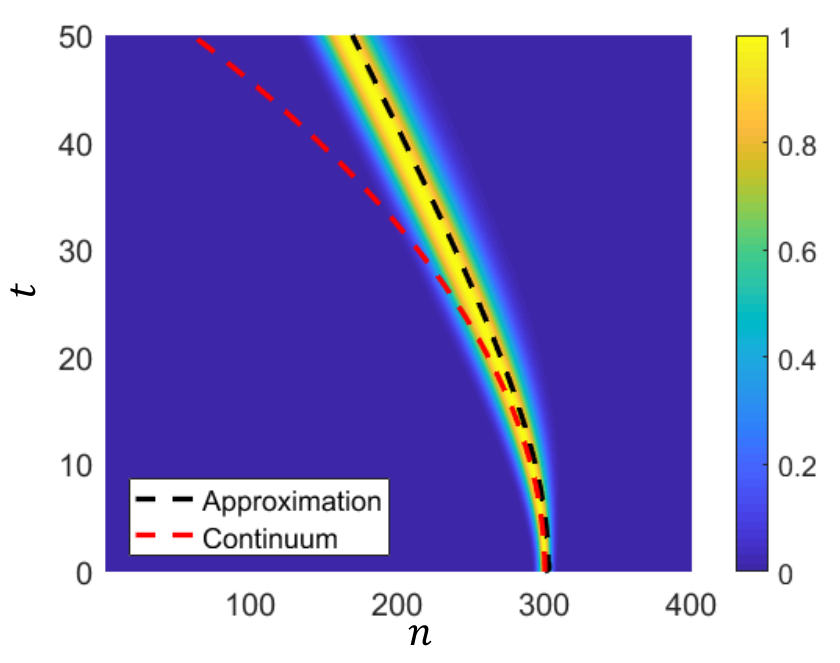}}\caption{\label{fig:gaussian_dynamics}Numerical results comparing the non-Hermitian
dynamics of a Gaussian wave-packet on the lattice versus the continuum
limit. While the initial acceleration is similar in both cases, only
the lattice dynamics eventually saturates to a maximum velocity.
See Video\_2 in the Supplementary Material for an animation of the dynamics.
The initial conditions are $n_{0}=300,$ $k_{0}=0$, $\sigma=3$,
the lattice spacing is $a=1$, and the parameter of the Hamiltonian
are $t_{l}=2,$ $t_{r}=1.5$. For the ease of presentation, the wave-function
is normalized such that $\max\left(|\psi_{n}|^{2}\right)=1$ for any
$t$. The black and red dotted-lines represent the approximation and
the continuum limit given by Eq. (\ref{eq:non_hermitian_approximation}),
(\ref{eq:max_position_continuum}) respectively.}
\end{figure}
In Fig. \ref{fig:gaussian_dynamics} we present numerical results
of the dynamics of a Gaussian wave-packet on the lattice and compare
the results to the approximation in Eq. (\ref{eq:non_hermitian_approximation}),
and the continuum limit solution in Eq. (\ref{eq:max_position_continuum}).
Initially, we can see that the lattice wave-packet has a similar acceleration
to that of the continuum. However, for large times, only the velocity
of the lattice wave-packet's peak saturates to a limiting velocity
$v_{\mathrm{nh}}$ (as we expect for any initial condition, since
there is an upper bound on the lattice velocity). We note that the
approximation, Eq. (\ref{eq:non_hermitian_approximation}), agrees
with the numerical results, although it shows a slight deviation in
the long-time limit. The reason for this is that our saddle-point approximation was constructed only in the regime $|v|<v_{\mathrm{h}}$, so it enforces $v_{\mathrm{h}}$ as the maximal velocity and cannot capture the true non-Hermitian front velocity $v_{\mathrm{nh}}$; see Appendix~\ref{appendix_Hermitian-dynamics-of-gaussian} for more details.

\subsubsection{\label{subsec:Non-Hermitian-Reflection}Non-Hermitian Reflection}

In the previous section we analyzed the dynamics of a Gaussian wave-packet
on the lattice, and saw that its velocity is bounded, unlike the continuum
case. We will now show that reflections from the edge can lead to
even more striking differences between the lattice and the continuum.
\begin{figure}
\centerline{\includegraphics[scale=0.5]{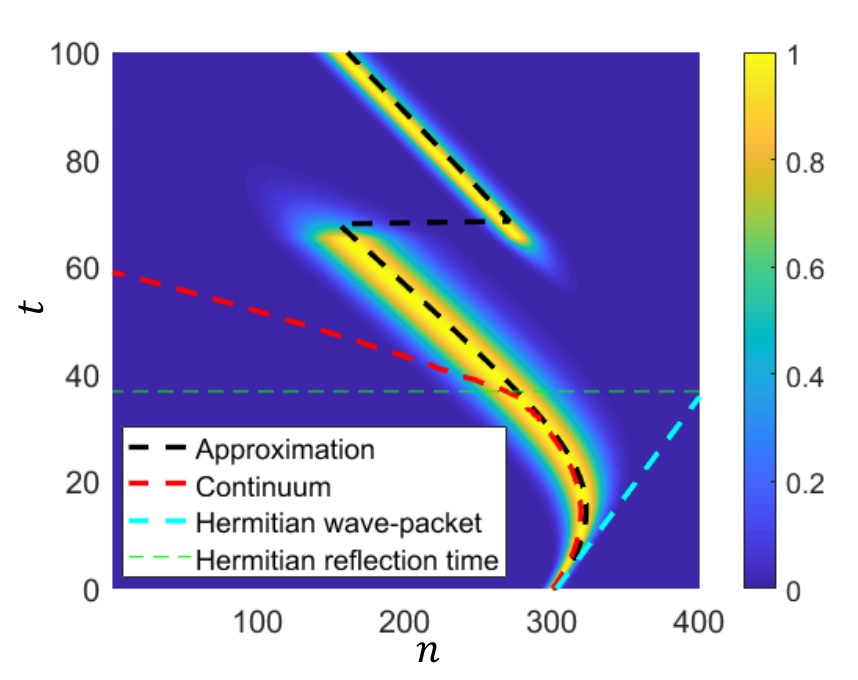}}\caption{\label{fig:gaussian_dynamics_reflection}Numerical results of a Gaussian
wave-packet, with initial parameters identical to those of Fig. \ref{fig:gaussian_dynamics},
except for $k_{0}$ which is now $\pi/4$ instead of $0$.
of propagation of the wave-packets). See Video\_3 in the Supplementary Material for an animation of the dynamics.
The behavior of the wave-packets in the continuum and on the lattice
is similar until the time when the Hermitian wavepacket (marked in
cyan) hits the wall: $t_{\mathrm{hit}}=\frac{100}{v_{0}}\approx37$
(marked in green), where $v_{0}=2\sqrt{t_{r}t_{l}}k_{0}$. The approximation
(marked in black) was calculated using the image method, and is in
agreement with the numerical results.}
\end{figure}
In Fig. \ref{fig:gaussian_dynamics_reflection} we can see a scenario
similar to that of Fig. \ref{fig:gaussian_dynamics}, with the only
difference being that $k_{0}=\pi/4$ instead of $0$, that is, the
initial velocity is to the right. We observe that the presence of
the wall causes a sudden change in the behavior of the wave-packet.
In the continuum, there is only a sharp change in the velocity. However,
on the lattice, the jump is in the \textit{position} of the wave-packet.

We will begin by analyzing the behavior in the continuum: At first,
we can see that the non-Hermitian wave-packet accelerates to the left.
It is already moving away from the wall, but then exhibits a sudden
change in behavior at $t_{\mathrm{hit}}=\frac{100}{v_{0}}\approx37$,
where $v_{0}=2\sqrt{t_{r}t_{l}}k_{0}$, which is precisely when the
Hermitian wave-packet hits the wall (see Appendix \ref{subsec:appendix_The-continuum-limit}).
The reason for this is that at this point the Hermitian wave-packet
changes its momentum from $k_{0}$ to $-k_{0}$, leading to an abrupt
change of $2k_{0}/m$ in the velocity of the \textit{non-Hermitian}
wave-packet (marked in red), since it is derived from the Hermitian
wave-packet \citep{Li2022} \textemdash{} see Eq. (\ref{eq:max_position_on_lattice}).
However, it is worth emphasizing that this is not the result of scattering
from the wall of the non-Herimitan wave-packet itself (as it never
even reached the wall), but rather its Hermitian component: In fact,
such scattering can occur regardless of how far the non-Hermitian
wave-packet is from the edge (as long as the Hermitian counterpart
hits the wall) \cite{inelastic_comment}.

We now turn to discuss the case of the lattice, where the reflection
process is somewhat more complex. Up to the time $t_{\mathrm{hit}}\approx37,$
the lattice and continuum systems exhibit similar behaviors. However,
while at $t_{\mathrm{hit}}$ the continuum system change its velocity
abruptly, on the lattice it remains the same until an peculiar event
happens at much lager time, around $t\approx70$. One may suspect
that this is related to the fact that the time when the Hermitian
wave-packet hits the wall is different on the lattice. But this is
certainly not the case, as this time is given by $t_{\mathrm{hit}}^{\mathrm{lattice}}=\frac{100}{2\sqrt{t_{r}t_{l}}\sin(k_{0})}\approx41$.
We also emphasize that this is not a numerical artifact of some sort,
as this result is supported by our approximate analytical solution
(see black dotted line). The actual explanation is tied to the asymmetry
of the wave-packet, arising from the velocity limit on the lattice.

Since there is an upper velocity bound on the lattice, any Gaussian
with a non-zero value for $k_{0}$ will not remain symmetric as it
evolves over time \citep{Schoenhammer2019}. Intuitively, this is
because the velocity range to the right of the wave-packet's peak
$[v_{0},v_{\max}],$ differs from the range $[-v_{\max},v_{0}]$ to
the left. This contrasts with the continuum case, where the Gaussian
remains symmetric for any $t$.
\begin{figure*}
\centerline{\includegraphics[scale=0.6]{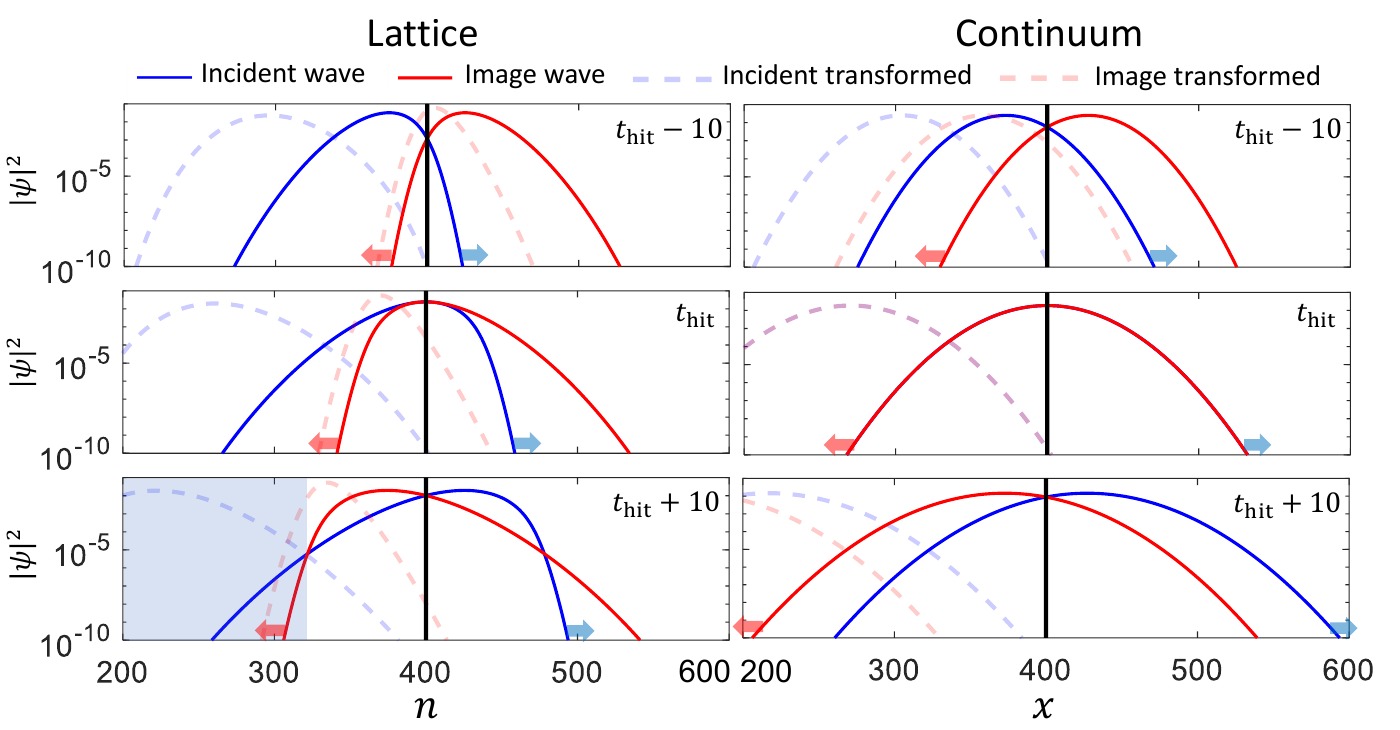}}\caption{\label{fig:gaussian_dynamics_reflection_hermitian}A Hermitian Gaussian
wave-packet (in blue) and its image (in red) moving towards a wall
at $n=400$, on the lattice (left column) and in the continuum (right
column). The rows represent the timestamps: $t_{\mathrm{hit}}-10$,
$t_{\mathrm{hit}}$, $t_{\mathrm{hit}}+10$ (top, middle, bottom,
respectively), where $t_{\mathrm{hit}}$ is the time where the packet
hits the wall ($t_{\mathrm{hit}}=\frac{d}{2\sqrt{t_{r}t_{l}}\sin(k_{0})}$
on the lattice and $t_{\mathrm{hit}}=\frac{d}{2\sqrt{t_{r}t_{l}}k_{0}}$
in continuum, where $d=100$). The dashed lines present the results
of the transformation (\ref{eq:Transformation}), normalized to 1,
which gives the non-Hermitian wave-packets that is derived from the
incident and image Hermitian counterpart, respectively.
See Video\_4 in the Supplementary Material for an animation in the case of the lattice.
The setup is the same as in Fig. \ref{fig:gaussian_dynamics_reflection_hermitian}
(note that $a=1$ is used, so that $n$ on the lattice is equivalent
to $x$ in continuum).}
\end{figure*}
This is demonstrated in Fig. \ref{fig:gaussian_dynamics_reflection_hermitian},
where we plot a wave-packet moving toward the wall (in blue), on the
lattice and in the continuum. To understand the mechanism of the reflection,
it is useful to consider also the image wave-packet (with respect to
the wall at $n=400$) which is presented in red (note that the actual
wave-packet hitting the wall, which is the sum of the incident and
the image wave-packets, is not depicted). The key point is that until the wave-packet
hits the wall ($t<t_{\mathrm{hit}}$) there is no significant difference
between the lattice and the continuum cases. However, at the time
of impact ($t=t_{\mathrm{hit}}$), in the continuum the incident and
the image wave-packets align precisely (being completely symmetric
with respect to their center), unlike the behavior on the lattice. A
direct consequence of this simple observation is that the non-Hermitian
counterparts (presented in dashed line in the figure) of the incident
and image wave-packets also align exactly at $t=t_{\mathrm{hit}}$
in the continuum\textbf{ }case (since they are obtained\textbf{ }using
a local transformation, the non-Hermitian wave-packets will align
if and only if the Hermitian wave-packets align). Indeed, this explain
the fact that $t=t_{\mathrm{hit}}$ is the exact moment when the non-Hermitian
wave-packet changes it velocity, as seen in Fig. \ref{fig:gaussian_dynamics_reflection_hermitian}.

However, on the lattice, a more subtle behavior occurs for $t>t_{\mathrm{hit}}$:
The asymmetry of the wave-packet leads to the formation of two distinct
regions. In one region (represented as the blurred area in the lower-left
panel of Fig. \ref{fig:gaussian_dynamics_reflection_hermitian}),
we continue to observe only the tail of the incident wave-packet that
has already struck the wall. In contrast, in the other region (represented
by the rest of the area up to the wall which is not blurred), we already
see the front of the image wave-packet which is actually the wave-packet
that was reflected from the wall. An alternative perspective is that,
due to the velocity limit on the lattice, the event of a ``wave-packet
hitting the wall'' does not propagate instantaneously to all of space,
in contrast with the continuum. Either way, the
two regions in the Hermitian wave-packet also correspond to two regions
in the non-Hermitian dynamics. In fact, as can be seen in the bottom-left
panel of the figure, in each region a different non-Hermitian wave-packet
is dominant. Therefore, we see now that the behavior that was observed
earlier in Fig. \ref{fig:gaussian_dynamics_reflection}, where the
wave-packet ``vanished'' at a certain point and reappeared somewhere
else, is actually just the right non-Herimitian wave-packet becoming
more dominant than the left one (and since the wave-function is normalized,
this means that the left wave-packet will be suppressed at this point
and become invisible). Actually, looking at this in log-scale we will
be able to see both of the wave-packets even for $t>80$; see the
accompanying animation for Fig. \ref{fig:gaussian_dynamics_reflection}
for more details.

We note that as the lattice spacing decreases, the ``transition time''
(that is, the time where the image wave-packet becomes more dominant)
approaches $t_{\mathrm{hit}}$, the time in continuum limit. In Appendix
\ref{subsec:appendix_non_hermitian_transition_time} we use our approximation
of Eq. (\ref{eq:non_hermitian_approximation}) to show in general
that for small $a$, this time is given by
\begin{equation}
t=t_{\mathrm{hit}}+t_{\delta},
\end{equation}
where 
\begin{equation}
t_{\mathrm{hit}}=\frac{d}{\frac{\sin(ak_{0})}{a}a_{0}^{2}\sqrt{t_{l,0}t_{r,0}}},
\end{equation}
is the Hermitian transition time, and the leading correction in terms
of $a$ is 
\begin{equation}
t_{\delta}\approx\frac{3}{64}\frac{\ln^{2}\left(\frac{t_{r,0}}{t_{l,0}}\right)}{\sqrt{t_{l,0}t_{r,0}}}\frac{\cos^{2}(ak_{0})}{\left(\sin(ak_{0})/a\right)^{3}}\frac{d^{3}}{a_{0}^{4}\sigma^{4}}a^{2},\label{eq:transition_time_correction}
\end{equation}
where $d$ is the initial distance from the wall, $a_0$ is a fixed reference lattice spacing used to define the continuum limit (we keep $a_0^2\sqrt{t_{l,0}t_{r,0}} = a^2\sqrt{t_lt_r}$ constant as $a \to 0$), and $t_{r,0},t_{l,0}$ are the right
and left hopping parameters corresponding to $a_{0}$ --- see Appendix \ref{subsec:appendix_The-continuum-limit}
for more details.

\subsubsection{Critical threshold $\sigma_{c,\mathrm{ref}}$}

It turns out that such an abrupt change in the wave-packet
position can only be observed its initial width $\sigma$ exceeds
a certain threshold. We will outline the main idea here, with further
details presented in Appendix \ref{subsec:The-effects-of_sigma_d_on_transition}.
To understand why this occurs, we can use reasoning similar to that
used for calculating the transition time in Eq. (\ref{eq:transition_time_correction}).
In the presence of a wall, there are two wave-packets: the original
and its image, induced by the wall. As discussed earlier, both wave-packets
initially accelerate to the left but with opposite quasi-momentum
$k_{0}$. The image wave-packet starts with an initial amplitude that
is lower by a factor of $r^{2d}$. As discussed above, the wave-packets
on the lattice do not accelerate indefinitely, but eventually reach
a finite velocity. When both wave-packets achieve this limiting velocity,
their amplitudes will continue to grow at the same rate, and their
ratio will remain constant from that point onward. Therefore, for
the image wave-packet to become more dominant than the real wave-packet,
it must outpace it early on. Whether this occurs is highly sensitive
to the initial conditions of $\sigma$ and the initial distance from
the wall $d$ due to the following reason: the smaller $\sigma$ is
in real space, the larger the non-Hermitial acceleration becomes (see
Eq. (\ref{eq:max_position_continuum})), causing the wave-packet to
quickly saturate at the limiting value. Another way to understand
this is that the smaller $\sigma$ is, the larger the initial condition
for the most significant momentum component ($-\pi/2$ in our case),
reducing the time it takes to reach the limiting velocity. For instance,
if $\sigma=0$ (a single-site starting condition), the wave-packet
begins with the final non-Hermitian velocity, and since the image
wave-packet also starts at this limiting velocity, it will never overtake
the original. Therefore $\sigma$ must exceed a certain value to allow
the image wave-packet enough time to catch up with the original. We
also note that a larger $d$ increases their initial difference (by $r^{2d}$), requiring $\sigma$ to be even larger. This behavior is also illustrated
in Fig. \ref{fig:appendix4_A} in Appendix \ref{subsec:The-effects-of_sigma_d_on_transition}.

\subsection{Effects of weak disorder\label{subsec:Disorder}}

In this section, we investigate the effect of weak disorder, which
does not localize the wave-packet on the scale of the system size.
Related work on ``non-Hermitian jumps''  considers strong disorder which leads to Anderson localization, and thus addresses a different parameter regime \citep{Weidemann2021,PhysRevResearch.3.013208,PhysRevB.106.064205,PhysRevResearch.4.043081,Longhi_2023,PhysRevA.110.053517,PhysRevB.111.214204,kokkinakis2025nonhermitianoffdiagonaldisorderedoptical}.
While weak disorder barely alters
the dynamics in Hermitian systems, we will show that in non-Hermitian
systems, it can lead to the emergence of a wave-packet, similar to
the effect seen earlier (caused by the reflection of the Hermitian
wave-packet from the wall). However, in this case, the cause is the
disorder. As we will discuss in detail below, the main reason for
this behavior is that in non-Hermitian systems, the tail of the wave-packet in position space (typically insignificant in Hermitian systems) dominates the long-time
dynamics due to the transformation in Eq. (\ref{eq:Transformation}).
From another perspective, weak disorder causes weak scattering to
all $k$-states, but even small occupations of certain $k$-states
will grow exponentially in non-Hermitian systems. This phenomenon
occurs only when the initial width of the wave-packet exceeds
a certain threshold, denoted as $\sigma_{c,\mathrm{dis}}$, which
depends on the disorder strength and differs from the previously discussed
critical value $\sigma_{c,\mathrm{ref}}$.

A typical way of introducing disorder is to consider an additional
on-site term to the Hamlitonian: $H'=\sum_{n}w_{n}\left|n\right\rangle \left\langle n\right|$,
where $w_{n}$ are uniformly distributed in $[-w,w]$.
\begin{figure}
\centerline{\includegraphics[scale=0.5]{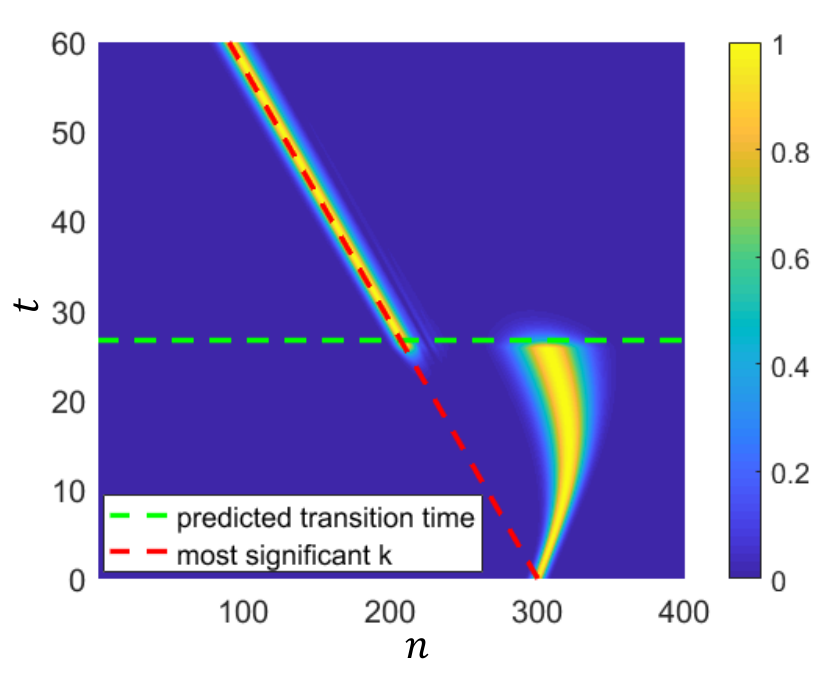}}\caption{\label{fig:disorder_transition}Numerical results of a Gaussian wave-packet,
with initial parameters identical to those of Fig. \ref{fig:gaussian_dynamics_reflection},
expect that the Hamiltonian also contain an onsite disorder realization
taken from a distribution with $w=10^{-7}$. Unlike Fig. \ref{fig:gaussian_dynamics_reflection},
where the transition was due to the existence of the wall, here it
is the result of the disorder.
See Video\_5 in the Supplementary Material for an animation of the dynamics.
The dashed green
line represents the predicted transition time $t\approx27$, calculated
as described in the main text. The dashed red line indicates the most
significant $k$ wave-packet, which becomes more prominent at the
transition time.}
\end{figure}
 In Fig. \ref{fig:disorder_transition} we present the dynamics of
a Gaussian wave-packet in the presence of a disorder realization with
$w=10^{-7}$, $N=400$ ($a=1$). We note that since $\xi\sim1/w^{2}\gg N$
(see Appendix \ref{sec:Localization-length-estimation}), localization
effects will not be observed. Similarly to the reflection case, a
transition from the initial wave-packet to another occurs at a certain
time. However, in the current case the cause is disorder rather than a wall
(here existence of the wall is irrelevant). Also, here the transition
happens sooner than in the no-disorder scenario. The main idea will
be introduced here while more technical details are provided in Appendix
\ref{subsec:Appendix_Disorder}.

Examining the wave-packet in $k$-space, the disorder enables transitions
from the center of the wave-packet (initially positioned at $k_{0}$)
to all other $k$ values. In the case of a Hermitian Hamiltonian,
this scattering process continues until it reaches a saturation value
determined by the disorder strength $w$ and the system size. However,
since the Hamiltonian is non-Hermitian, each $k$-value also experiences
exponential growth or decay. In particular, the most significant $k$-value
which is $k=-\pi/2$ (since it exhibits the fastest growth, as was
discussed earlier), will eventually surpass the initial Gaussian wave-packet
in magnitude. This explains the transition presented in the figure.

To estimate the transition time, we develop approximations for
the amplitudes of both the $k=-\pi/2$ packet and the initial Gaussian wave-packet. We start with the $k=-\pi/2$ packet. Using perturbation theory and averaging over the disorder realizations (denoted by $\left\llbracket \right\rrbracket$), we find that at small times
\begin{equation}
\left\llbracket |c_{k}(t)|^{2}\right\rrbracket-|c_{k}(0)|^{2} \approx \frac{t^{2}}{\hbar^{2}}\frac{w^{2}}{3}\frac{1}{N}.
\end{equation}
At intermediate times this increase saturates, with saturation value and saturation time given approximately by
\begin{equation}
V_{s}\approx\frac{w^{2}}{3}\frac{1}{N}\frac{1}{\mathrm{Re}\left(E_{k}-E_{k_{0}}\right)^{2}},\,t_{s}\approx\frac{\pi}{\mathrm{Re}(E_{k}-E_{k_{0}})},\label{eq:saturation_value_k}
\end{equation} and at later times the non-Hermitian exponential growth sets in:
\begin{equation}
|\psi(k,t)|^{2}\approx V_{s}e^{2(t_{l}-t_{r})(t-t_{s})}.\label{eq:_most_significant_k_amplitude}
\end{equation}
Plugging
in $k_{0}=\frac{\pi}{4}$ and $k=-\pi/2$, yields our first approximation.

For the second approximation, as we observed above, up to intermediate times the Gaussian is well described by the ``continuum limit'' Hamiltonian. We thus obtain
\begin{equation}
|\psi(k_{m},t)|^{2}\approx\left(\frac{8\pi\sigma^{2}}{N^{2}}\right)^{1/2}e^{2k_{0}bt+\frac{b^{2}}{2\sigma^{2}}t^{2}},\label{eq:maxmim_gaussian_amplitude_continuum}
\end{equation}
where $k_{m}$ is the maximum of the wave-packet, and $b=\ln\left(\frac{t_{r}}{t_{l}}\right)\sqrt{t_{l}t_{r}}$. We have verified that both of these approximations are in a very good agreement
with the numerical results (see Fig. \ref{fig:appendix_disorder_predictions} in Appendix \ref{subsec:Appendix_Disorder}).

By comparing Eq. (\ref{eq:maxmim_gaussian_amplitude_continuum})
and (\ref{eq:_most_significant_k_amplitude}), we obtain a quadratic
equation, and its solution readily gives the transition time, marked by the green dashed line in Fig. \ref{fig:disorder_transition}. Finally, it is worth noting that
even if no disorder is considered ($w=0$), numerical artifacts may
induce such transition. This is because the numerical accuracy of
the computer may not be sufficient to capture the evolution of the tail
of the wave-packet, effectively introducing a disorder term with amplitude
related to the computer precision threshold. To overcome this (for
example, to plot the disorder-less scenario that was presented in
Fig. \ref{fig:gaussian_dynamics_reflection}) one may need to increase
the precision by using an arbitrary precision library (such as Advanpix
\citep{Advanpix} in Matlab).

\begin{figure}
\centerline{\includegraphics[scale=0.5]{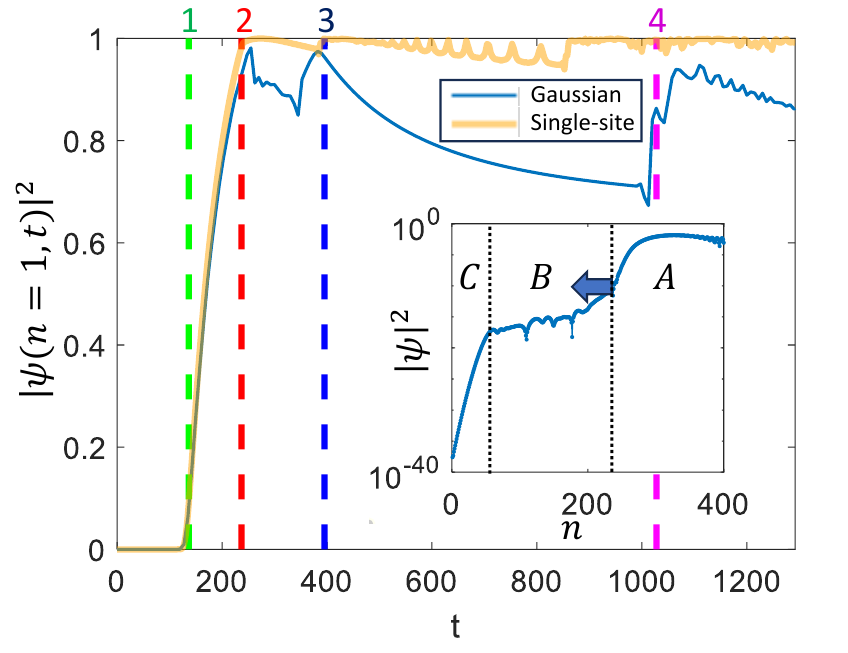}}\caption{\label{fig:disorder_dynamics_of_first_site}Blue line: the numerical
results of the amplitude at the leftmost site as a function of time,
where the initial conditions are the same as in Fig. \ref{fig:disorder_transition}
(a Gaussian with initial position $x_{0}=\frac{3N}{2}$). Orange line:
the same, but for initial condition $\psi(n,t=0)=\delta_{n,x_{0}}$
(for $t>t_{2}$ only the envelope of the wave-packet is shown). We
mark the following timestamps (in green, red, blue, and magenta, respectively):
$t_{1}=\frac{x_{0}}{v_{\mathrm{nh}}}$, $t_{2}=\frac{x_{0}}{v_{\mathrm{h}}},$
$t_{3}=\frac{2N-x_{0}}{v_{\mathrm{h}}}$, $t_{4}=\frac{4N-x_{0}}{4}\frac{1}{v_{\mathrm{h}}}$,
where $v_{\mathrm{h}},$$v_{\mathrm{nh}}$ are the Hermitian and non-Hermitian
velocities, given by Eq. (\ref{eq:Hermitian_velocity}),(\ref{eq:nonHermitian_velocity})
respectively. Inset: the Hermitian counterpart of the wave-packet,
as a function of $n$, for $t=190$ (which is roughly $\frac{t_{1}+t_{2}}{2}$).
The boundary between regions A and B is given by $2N-x_{0}-v_{\mathrm{h}}t,$
while the boundary between region B and C is given by $x_{0}-v_{\mathrm{h}}t$
.}
\end{figure}

We can gain more insights by looking at the amplitude of the leftmost
site (the left boundary), see Fig. \ref{fig:disorder_dynamics_of_first_site}.
As was demonstrated earlier, for a better understanding of the dynamics
it is best to start by examining the behavior of the Hermitian counterpart
of the wave-packet: In the inset, we present the Hermitian wave-packet
(initially moving to the right) at a time when it has already struck
the right wall and now advancing toward the left wall. We highlight
three distinct regions: (A) The Gaussian wave-packet itself. (B) The
cutoff of the Gaussian caused by disorder: Instead of the expected
Gaussian decrease in the absence of disorder, we observe a saturation
in the value of $|\psi|^{2}$. Its specific features depends on the
particular disorder realization, while its amplitude is determined
by $w$ (see Eq. \ref{eq:saturation_value_k}). (C) The tail of the
distribution, that is, the wave-packet beyond the maximum Hermitian
velocity $x_{0}-v_{\mathrm{h}}t$.

We move on to analyzing the non-Hermitian dynamics. In blue we present
the result for the Gaussian initial condition (that was discussed
here) while for comparison, we add in orange the result for the initial
condition $\delta_{n,x_{0}}$ (as was discussed at section \ref{sec:single_site_starting_condition_dynamics}).
We mark four timestamps: the first two, $t_{1}=\frac{x_{0}}{v_{\mathrm{nh}}},$
$t_{2}=\frac{x_{0}}{v_{\mathrm{h}}}$, represent the time that takes
for the Hermitian and non-Hermitian wave-packets to reach the boundary, respectively
(where the initial expectation value of the distribution is $x_{0}=\frac{3N}{4}$).
First, we note that up to $t=t_{2}$, the behavior of the two initial
conditions (blue and orange lines) is remarkably similar (see also
Fig. \ref{fig:delta_dynamics_of_first_site}). This fact implies that
the behavior of any (localized) initial condition with some disorder
would also look similar up to this time. The reason for this is that
in the presence of disorder, the tail of the Hermitian wave-packet
(i.e., the wave-packet beyond position $x_{0}-v_{\mathrm{h}}t$, corresponding
to region C in the Hermitian case) will look similar for any initial
distribution, and the tail is the important factor controlling the
properties of the non-Hermitian wave-packet.

As for $t_{2}$, this is the time where the disorder-dependent part
of the Hermitian wave-packet (that was marked as region B in the inset
of Fig. \ref{fig:disorder_dynamics_of_first_site}) hits the left
wall. Then, at $t=t_{3}$, the front of the Gaussian wave-packet itself
(region A) also arrives to the left boundary. Finally, $t_{4}$ is
the time where the Gaussian wave-packet hits the left boundary for
the second time (after being reflected from the other end of the system)\footnote{We note that the single-site initial condition (orange line) exhibit
some change at some time $t_{3}<t<t_{4}$. This time equals to $\frac{2N+x_{0}}{v_{h}}$,
which is the time it takes for the left-moving part of the wave-packet
to hit the left wall, then the right wall, and then the left wall
again. This part is suppressed in the Gaussian dynamics, since the
Hermitian Gaussian is only moving to one direction, in contrast with
the single-site initial condition, which generates two symmetrical
wave-packets moving to the opposite directions.}. This once again demonstrates that only the Hermitian wave-packets
can be reflected.

\subsubsection{Critical threshold $\sigma_{c,\mathrm{dis}}$}

The disorder-induced transition discussed above can only be observed
when the initial width exceeds a critical value, $\sigma_{c,\mathrm{dis}}$,
which depends on the disorder strength $w$. To understand this, it
is useful to revisit the wave-packet in quasi-momentum space. As noted
earlier, the saturation value of the most significant $q$ is proportional
to $w^{2}$, see Eq. (\ref{eq:saturation_value_k}). However, if $\sigma$
is sufficiently small in real space, it may be large enough in $q$-space
to surpass this saturation value. When this occurs, the effect of
disorder at $q=-\pi/2$ becomes negligible, resulting in a smooth
transition. This behavior is illustrated in Fig. \ref{fig:appendix_disorder_critical_value}
in Appendix \ref{subsec:Appendix_Disorder}. The critical value $\sigma_{c,\mathrm{dis}}$
can be explicitly determined by comparing
\begin{equation}
\left|\psi(q=-\frac{\pi}{2})\right|^{2}=\left(8\pi\frac{\sigma^{2}}{N^{2}}\right)^{1/2}e^{-2\frac{\sigma^{2}}{a^{2}}(\frac{\pi}{2}+q_{0})^{2}}
\end{equation}
to the saturation value from Eq. (\ref{eq:saturation_value_k}) and
extracting $\sigma$. Finally, we note that for the same reasons,
given a fixed value of $\sigma,$ the transition can only be observed
when the disorder exceeds a certain critical value $w_{c}$, which
depends on $\sigma.$

\section{\label{sec:Beyond-the-Hatano-Nelson}Beyond the Hatano-Nelson model}

\begin{figure}
\centerline{\includegraphics[scale=0.5]{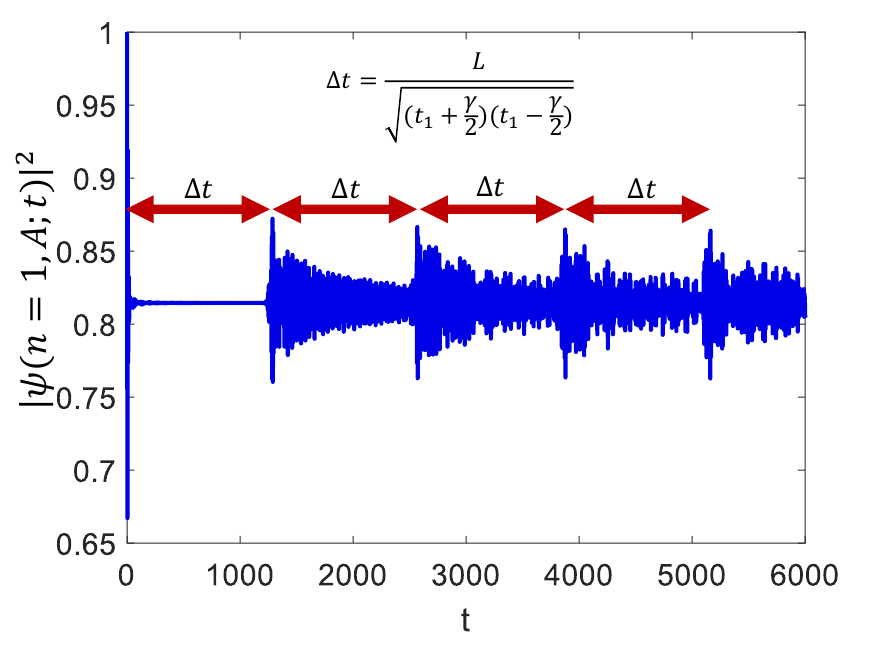}}\caption{\label{fig:SSH_non_hermitian}Numerical calculation of the amplitude
at the leftmost site as a function of time, for the non-Hermitian
SSH model in Eq. (\ref{eq:non-Hermitian_SSH}). The parameters are
$t_{1}=0.4,$ $t_{2}=1$, $\gamma=0.5$. The wavefunction is initially
localized at the leftmost site. The time difference between each peak
is given by $\Delta t=2L/v_{h}$, where $v_{h}=2\sqrt{(t_{1}+\gamma/2)(t_{1}-\gamma/2)}$
is the velocity of the Hermitian wave-packet.}
\end{figure}

While we have focused on the Hatano-Nelson model for simplicity, most
of the discussion above\textemdash specifically the coexistence of
Hermitian and non-Hermitian wave-packets\textemdash should also apply
to more complex models, provided that the Hamiltonian is pseudo-Hermitian.
For example, we will now focus on a specific version of the non-Hermitian
SSH model \citep{Lieu2018,Kunst2018}. We consider the Hamiltonian
\begin{gather}
H=\sum_{n}\left(t_{1}-\frac{\gamma}{2}\right)\left|n,A\right\rangle \left\langle n,B\right|+\left(t_{1}+\frac{\gamma}{2}\right)\left|n,B\right\rangle \left\langle n,A\right|\label{eq:non-Hermitian_SSH}\\
+\left(t_{2}\left|n,B\right\rangle \left\langle n+1,A\right|+h.c.\right),\nonumber 
\end{gather}
which is pseudo-Hermitian, as it can become Hermitian using the transformation
$H'=S^{-1}HS$ where
\begin{gather}
S=\left|1,A\right\rangle \left\langle 1,A\right|+r^{N}\left|N,B\right\rangle \left\langle N,B\right|\nonumber \\
+\sum_{n=1}^{N-1}\left[r^{n}\left|n,B\right\rangle \left\langle n,B\right|+r^{n}\left|n+1,A\right\rangle \left\langle n+1,A\right|\right]\label{eq:non_hermitian_SSH_transformation}
\end{gather}
(that is, the diagonal components are $1,r,r,r^{2},r^{2},...,r^{N-1},r^{N-1},r^{N}$, where $r = \sqrt{(t_{1}+\gamma/2)/(t_{1}-\gamma/2)}$).
We emphasize again that the discussion in Sec. \ref{sec:Deriving-the-dynamics}
remains valid, meaning that we anticipate the coexistence of Hermitian
and non-Hermitian wave-packets in this setup as well. Consider, for
example, the configuration in Fig. \ref{fig:SSH_non_hermitian}, which
illustrates the dynamics of the leftmost site. Initially, the wavefunction
is localized at $n=1,$ but as it evolves, the \textit{Hermitian}
component of the wave-packet propagates to the other edge of the system
and gets reflected back. Hence, we observe oscillations with a period of $\Delta t=2L/v_{h}$,
where $v_{h}=2a\min(\tilde{t}_{1},t_{2})$ is the velocity of the
Hermitian SSH model and $\tilde{t}_{1}=\sqrt{(t_{1}+\gamma/2)(t_{1}-\gamma/2)}$
is the effective hopping amplitude after the transformation.
Notably, the saturation of the average amplitude at approximately
0.8 suggests the presence of a topological edge state, indicating
that classifying the topological phase based on dynamics may be possible.
We plan to examine this in more detail in future work. Finally, we
note that in principle, all of the other phenomena discussed in Sec.
\ref{sec:Non-Hermitian-wave-packet-dynami} can also be observed in
this setup, but in practice some of them would be more elusive. We
elaborate on this in Appendix \ref{subsec:Appendix_beyond_HN}.

\section{Discussion and Conclusions}

In this paper we explored the dynamics of pseudo-Hermitian systems,
highlighting the unexpected coexistence of Hermitian and non-Hermitian
wave-packets. Focusing primarily on the Hatano-Nelson model with open
boundary conditions, we demonstrated that the two dynamics are interconnected
through a local transformation. Importantly, we show that the Hermitian
wave-packet is not merely a mathematical abstraction; it exists within
the system, propagates, and can reflect off boundaries. This is evident
even in the simplest case of a single-site initial condition (i.e.,
when the width $\sigma\ll a$), where oscillations near
the boundaries occur with a period matching the time taken for the
Hermitian wave-packet to travel to the opposite end and return.

For moderate widths (i.e., $a<\sigma\ll L$), we showed
that the coexistence of these wave-packets is essential for explaining
more complex phenomena, such as non-Hermitian reflection at the boundary.
The transition occurs when the Hermitian wave-packet hits the wall
(although it may not be observed until some time later due to lattice
effects). Notably, the non-Hermitian wave-packet does not need to
be close to the wall during the transition, reinforcing the idea that
the Hermitian wave-packet is the crucial element in the reflection
process. Additionally, we revealed that the transition due to disorder,
typically attributed to the amplification of certain $k$-values,
can also be interpreted as the tail of the Hermitian wave-packet becoming
dominant over time due to the exponential transformation connecting
the two dynamics. Finally, we examine a version of
the non-Hermitian SSH model to demonstrate that our results generalize
to pseudo-Hermitian models beyond just the Hatano-Nelson model. We
believe that this new perspective on the dynamics can significantly
enhance the understanding of pseudo-Hermitian systems.

A pertinent question arises: Can the tool of investigate the dynamics
using a transformation remain effective when the Hamiltonian is not
pseudo-Hermitian? Preliminary results for simpler models, such as
the Hatano-Nelson model with complex coefficients, suggest that this
approach is still beneficial. By applying a transformation, we can
attain a Hamiltonian that, while not being Hermitian, does not favor
one direction (i.e., $|t_{l}|=|t_{r}|$), which can simplify the analysis.
However, for more complex models, such as the general non-Hermitian
SSH model, it remains unclear to what extent the methodology developed
here will apply. Future research will delve deeper into cases beyond
the pseudo-Hermitian framework.

Another promising direction for investigation is to leverage the tools
derived here to explore the topology of pseudo-Hermitian systems and
classify it based on their dynamics. For this, we can consider pseudo-Hermitian
topological models such as the non-Hermitian SSH model discussed in
Sec. \ref{sec:Beyond-the-Hatano-Nelson}. These tools can assist in
investigating how non-Hermitian effects influence topological properties,
particularly the stability and behavior of edge states.

\textbf{Note added.} After completing this work we have become aware of the very recent preprint by He and Ozawa (Ref. \citep{he2025}) on wave-packet dynamics in the Hatano-Nelson model, reporting a similar boundary-associated non-Hermitian reflection (jump). However, their analysis emphasizes the point of view of temporal Goos--H\"anchen shifts at the edge, while we uncover the coexistence of the Hermitian and non-Hermitian wave-fronts as the basis for this phenomenon. 

\subsection*{ACKNOWLEDGMENTS}

We would like to thank Guy Gabrieli for unpublished results which motivated this study. Our work has been supported by the Israel Science Foundation (ISF) and the Directorate for Defense Research and Development (DDR\&D) Grant No. 3427/21, the ISF Grant No. 1113/23, and the US-Israel Binational Science Foundation (BSF) Grants No. 2020072 and 2024140.

\appendix

\section{\label{appendix_Hermitian-dynamics-of-gaussian}Hermitian dynamics
of a Gaussian wave-packet}

Suppose we have a Gaussian wave-packet, given by
\begin{equation}
\left|\psi(t=0)\right\rangle =N\sum_{n}e^{-\frac{n^{2}a^{2}}{4\sigma^{2}}+ik_{0}an}\left|n\right\rangle ,\label{eq:appendix_starting_condition}
\end{equation}
where $N\approx(2\pi\sigma^{2}/a^{2})^{-\frac{1}{4}}$. We will now
compute $\psi(m,t)=\left\langle m\left|e^{-iHt}\right|\psi(0)\right\rangle $,
where the Hermitian Hamiltonian is given by
\begin{equation}
H=t_{0}\sum_{n}\left|n\right\rangle \left\langle n+1\right|+h.c.\label{eq:appendix_Hermitian_Hamiltonian}
\end{equation}
We first notice that the expression can be written as the discrete
convolution of $\psi$ and the propagator, that is
\begin{equation}
\psi(m,t)=\sum_{n}g(m-n,t)\psi_{0}(n),\label{eq:appendix_convolution}
\end{equation}
where we have defined $g(m-n,t)\equiv\left\langle m\left|e^{-iH't}\right|n\right\rangle $
(due to discrete transitional invariance $g$ is only a function of
only the difference $m-n$, see Eq. (\ref{eq:hermitian_propagator}))
and 
\begin{equation}
\psi_{0}(n)\equiv\left\langle n|\psi(0)\right\rangle =Ne^{-\frac{n^{2}a^{2}}{4\sigma^{2}}+ik_{0}an}.
\end{equation}

For large $N$, we can use discrete Fourier transform. For any function
$f:\mathbb{N}\rightarrow\mathbb{R}$ we can define $F:(-\pi,\pi)\rightarrow\mathbb{R}$
by the relations
\begin{equation}
f(n)=\frac{1}{2\pi}\int_{-\pi}^{\pi}F(q)e^{iqn}dq,\,F(q)=\sum_{n=-\infty}^{\infty}f(n)e^{-iqn},
\end{equation}
where we define $q=ka$. Taking the limits of the sum in Eq. (\ref{eq:appendix_convolution})
to infinity, and using the convolution theorem, we get that
\begin{equation}
\Psi(q,t)=G(q,t)\Psi_{0}(q),\label{eq:appendix_convolution_theorem}
\end{equation}
where $\Psi(q,t),G(q,t),\Psi_{0}(q)$ are the discrete Fourier transform
of $\psi(n,t),g(n,t),\psi_{0}(n)$, respectively, and we note that
$G(q,t)=e^{-2it_{0}\cos(q)t}.$ We can calculate $\Psi_{0}(q)$ explicitly:
Employing the fact that
\begin{equation}
\sqrt{\frac{\pi}{\sigma^{2}}}e^{-\frac{n^{2}}{4\sigma^{2}}}=\int_{-\infty}^{\infty}e^{-inx}e^{-\sigma^{2}x^{2}}dx,
\end{equation}
and using the Poisson summation formula
\begin{equation}
\frac{1}{2\pi}\sum_{n}e^{int}=\sum_{m}\delta(t-2\pi T),
\end{equation}
we obtain
\begin{equation}
\Psi_{0}(q)=\sum_{m}e^{-\frac{\sigma^{2}}{a^{2}}\left(q-q_{0}-2\pi m\right)^{2}}.
\end{equation}
Assuming also that $\sigma\apprge a$ (and $|q_{0}|\le\pi/2$), the
only significant contribution comes from $m=0$ so we are left with
$\Psi_{0}(q)=e^{-\frac{\sigma^{2}}{a^{2}}(q-q_{0})^{2}}.$ Performing
the inverse transform on Eq. (\ref{eq:appendix_convolution_theorem}),
we get
\begin{equation}
\psi(m,t)=N_{q}\int_{-\pi}^{\pi}e^{-\frac{\sigma^{2}}{a^{2}}(q-q_{0})^{2}}e^{-2it_{0}\cos(q)t}e^{iqm}dq,\label{eq:A11}
\end{equation}
\begin{equation}
N_{q}=\frac{(8\pi\sigma^{2}/a^{2})^{1/4}}{2\pi}.
\end{equation}

\begin{figure}
\centerline{\includegraphics[scale=0.5]{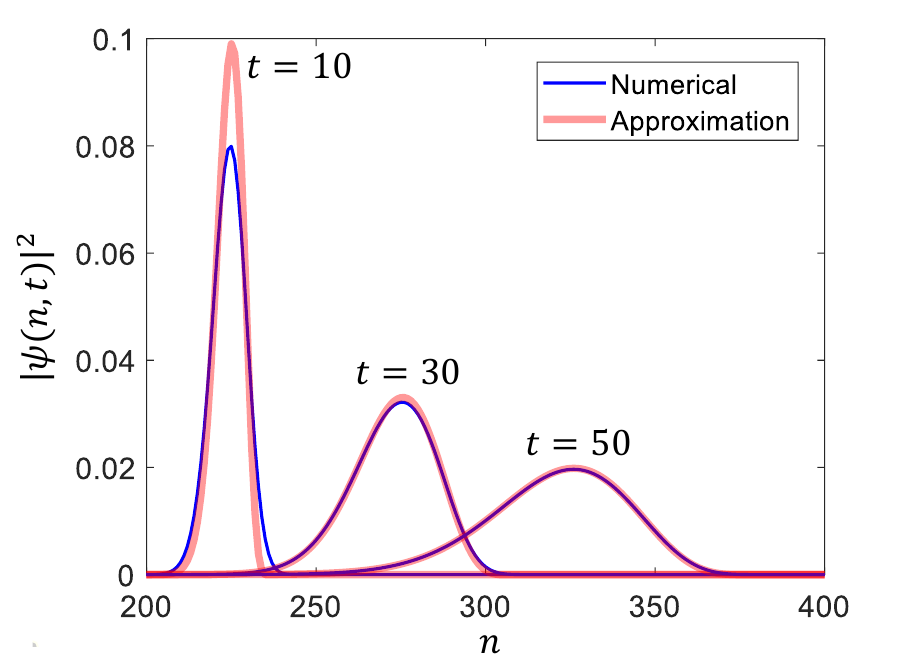}}

\caption{\label{fig:appendix_test_approximation}Numerical calculation of $e^{-iHt}\left|\psi_{0}\right\rangle ,$
where $H$ is given by Eq. (\ref{eq:appendix_Hermitian_Hamiltonian})
and $\psi_{0}$ is the Gaussian given by Eq. (\ref{eq:appendix_starting_condition}),
versus the approximation given by Eq. (\ref{eq:appendix_final_approximation}),
at times $t=10,30,50$. As $t$ grows, the Gaussian wave-packet acquires
a non-symmetrical shape, due to the fact that the maximum velocity
on the lattice is finite. The parameters are $n_{0}=200$, $\sigma=3,$
$a=1$, $k_{0}=\pi/4$, and $t_{0}=\sqrt{3}$. See Video\_6 in the Supplemental Material for an animation of the dynamics.} 
\end{figure}

To proceed, it will be convenient to define a ``velocity'' $v(m,t)$
by $vt\equiv am$. Then, we can rewrite Eq. (\ref{eq:A11}) as
\begin{equation}
\psi(v,t)=N_{q}\int_{-\pi}^{\pi}e^{-\frac{\sigma^{2}}{a^{2}}(q-q_{0})^{2}}e^{i\frac{tv_{\max}}{a}A(q,v)}dq,
\end{equation}
where we have defined
\begin{equation}
A(q,v)\equiv\frac{v}{v_{\max}}q-\cos(q).\label{eq:appendix_A_saddle_point}
\end{equation}
For $t\gg1/t_{0}$, this integral can be evaluated by the stationary
phase approximation:
\begin{multline*}
\psi(v,t)\approx N_{q}\sum_{j}e^{i\frac{tv_{\max}}{a}A(q_{j},v)}e^{-\frac{\sigma^{2}}{a^{2}}(q_{j}-q_{0})^{2}}\times\\
\times\sqrt{\frac{2\pi}{\frac{tv_{\max}}{a}\left|\frac{\partial^{2}}{\partial^{2}q}A(q_{j},v)\right|}}e^{\mathrm{\mathrm{sign}\left(\frac{\partial^{2}}{\partial^{2}q}A(q_{j},v)\right)}\frac{i\pi}{4}},
\end{multline*}
where the sum $\sum_{j}$ is over all of the $q_{j}$ satisfying 
\begin{equation}
\frac{\partial}{\partial q}A(q_{j})=\frac{v}{v_{\max}}+\sin(q)=0.
\end{equation}
We will now note that for $|v|<v_{\max}$ the solution is real, and
complex otherwise. We will start with the case of $|v|<v_{\max}$
(where $v_{\max}\equiv2at_{0}$, as the Hermitian maximal velocity $v_h$ in the main text). In this case we obtain two solutions
\begin{equation}
q_{1}=\arcsin\left(-v/v_{\max}\right),\,q_{2}=\pi-q_{1},
\end{equation}
and therefore
\begin{equation}
A(q_{1,2},v)\equiv\frac{v}{v_{\max}}q_{1,2}\mp\sqrt{1-\left(v/v_{\max}\right)^{2}},
\end{equation}
\begin{equation}
\frac{\partial^{2}}{\partial^{2}q}A(q_{j},v)=\pm\sqrt{1-\left(v/v_{\max}\right)^{2}}.
\end{equation}
Without loss of generality, we assume that $|q_{1}-q_{0}|<|q_{2}-q_{0}|$
so that $e^{-\frac{\sigma^{2}}{a^{2}}(q_{1}-q_{0})^{2}}\ll e^{-\frac{\sigma^{2}}{a^{2}}(q_{2}-q_{0})^{2}}$
(since $\sigma\apprge a$), so we can keep only the first term and
finally arrive at
\begin{equation}
\psi(m,t)\approx\frac{\left(8\pi\sigma^{2}/a^{2}\right)^{1/4}}{\sqrt{2\pi}}e^{\frac{i\pi}{4}}e^{i\frac{tv_{\max}}{a}A}\frac{e^{-\frac{\sigma^{2}}{a^{2}}(\arcsin\left(-\frac{am}{tv_{\max}}\right)-q_{0})^{2}}}{\sqrt{\frac{t}{a}v_{\max}\sqrt{1-\left(\frac{am}{tv_{\max}}\right)^{2}}}},\label{eq:appendix_final_approximation}
\end{equation}
where 
\begin{equation}
A=\frac{v}{v_{\max}}\arcsin\left(-\frac{am}{tv_{\max}}\right)-\sqrt{1-\left(\frac{am}{tv_{\max}}\right)^{2}}.
\end{equation}
A comparison of this result to numerical results is shown in Fig.
\ref{fig:appendix_test_approximation}.

Except for when $am\approx tv_{\max}$ , we may do one more simplification
by setting $\arcsin\left(-v/v_{\max}\right)=q_{0}$ outside of the
exponent. Taking also the absolute value and switching back to $m,k$,
we finally get
\begin{multline}
\psi(m,t)\approx\frac{\left(8\pi\sigma^{2}/a^{2}\right)^{1/4}}{\sqrt{2\pi\frac{t}{a}v_{\max}\cos(ak_{0})}}\times\\
e^{\frac{i\pi}{4}}e^{i\frac{tv_{\max}}{a}A}e^{-\frac{\sigma^{2}}{a^{2}}\left(\arcsin\left(-\frac{am}{tv_{\max}}\right)-ak_{0}\right)^{2}}.\label{eq:eq:appendix_final_approximation_simpler}
\end{multline}
One can check that for $a\rightarrow0$ this results approaches the
known result in continuum (a Gaussian wave-packet moving at a constant
velocity while becoming broader).

As was shown in Fig. \ref{fig:appendix_test_approximation}, the results
for the case $|v|<v_{\max}$ provide an excellent approximation for
the Hermitian wave-packet when the time is sufficiently large. However,
it is important to note that for non-Hermitian wave-packets in the
long-time limit, the behavior is primarily governed by the tail of
the distribution, where the approximation breaks down. This implies
that, for accurate long-time approximations, considering the $|v|>v_{\max}$
case may also be necessary. Although for most phenomena presented
in this paper this additional consideration is not essential, for
completeness, we will also derive the approximation for $|v|>v_{\max}$.
We are looking for the solutions of $\frac{v}{v_{\max}}+\sin(q)=0,$which
will now be on the complex plane (and therefore we can use the method
of steepest decent). Plugging $q=q'+iq'',$ we are looking for the
solutions of
\begin{equation}
\frac{v}{v_{\max}}+\sin(q')\cosh(q'')+i\sinh(q'')\cos(q')=0.
\end{equation}
Since $q''=0$ will not yield a solution for $|v|>v_{\max}$, we take
$q''\neq0$ and get
\begin{equation}
q'=-\mathrm{sign}(v)\frac{\pi}{2},\;q''=-\mathrm{sign}(v)\mathrm{arccosh}\left(\frac{|v|}{v_{\max}}\right),
\end{equation}
where we chose the sign of $q''$ so that the critical point would
be a maximum, on a contour that passes through it with fixed $q'=-\mathrm{sign}(v)\frac{\pi}{2}$
(that is, on this contour the imaginary part in the argument of the
exponent is constant). Then, we can use the saddle point approximation
to find
\begin{gather*}
\psi(m,t)=\sqrt{2\pi}N_{q}e^{-i\frac{tv}{a}\mathrm{sign}(m)\frac{\pi}{2}}\times\\
e^{i\frac{\sigma^{2}}{a^{2}}2(-\mathrm{sign}(m)\frac{\pi}{2}-q_{0})\mathrm{sign}(m)\mathrm{arccosh}\left(\frac{|m|a}{v_{\max}t}\right)}\times\\
e^{\frac{\sigma^{2}}{a^{2}}\left[\mathrm{arccosh}^{2}\left(\frac{|m|a}{v_{\max}t}\right)-(-\mathrm{sign}(m)\frac{\pi}{2}-q_{0})^{2}\right]}\times\\
\frac{e^{-|m|\mathrm{arccosh}\left(\frac{|m|a}{v_{\max}t}\right)+\frac{tv_{\max}}{a}\sqrt{m^{2}-\frac{t^{2}v_{\max}^{2}}{a^{2}}}}}{\sqrt{m^{2}-\frac{t^{2}v_{\max}^{2}}{a^{2}}}}.
\end{gather*}

\section{\label{appendix_sbusec_small_and_large_t}Small $t$ expansion for
Eq. (\ref{eq:non_hermitian_approximation})}

As described by Eq. (\ref{eq:non_hermitian_approximation}) in the
main text, the evolved Gaussian wave-packet governed by the non-Hermitian
counterpart is approximately equal to
\begin{equation}
\left\langle m|e^{-iHt}|\psi(0)\right\rangle \approx\tilde{C}r^{m}e^{-\frac{\sigma^{2}}{a^{2}}\left(\arcsin\left(\frac{am}{tv_{h}}\right)-ak_{0}\right)^{2}},\label{eq:approximation_appendix}
\end{equation}
where $\tilde{C}$ is some constant, and we assume that $\tilde{n}_{0}=0$
for convenience. As shown in Fig. \ref{fig:gaussian_dynamics} in
the main text, this approximation is in a good agreement with the
numerical results except for two discrepancies. The first one is just
a small shift of $\frac{2\sigma^{2}}{a^{2}}\ln r$ in its initial
position (see Eq. (\ref{eq:shifted_expectation_value})), which is
due to the fact that at $t=0$ the approximation is localized at a
certain position (since it has no width, the transformation $r^{m}$
cannot move its center ``back''). The second one is just the fact that the velocity of the approximation does saturates to $v_{\mathrm{h}}$ instead of $v_{\mathrm{nh}}$, as discussed in the main text.

We begin by finding the peak of $\left|\left\langle m|e^{-iHt}|\psi(0)\right\rangle \right|^{2}$
by maximizing the parameter of the exponent. That is, $m_{\max}$
is obtained from a solution of the equation
\begin{multline}
2(\sigma/a)^{2}\left(\arcsin\left(\frac{a}{tv_{h}}m\right)-ak_{0}\right)\frac{\frac{a}{tv_{h}}}{\sqrt{1-\left(\frac{am}{tv_{h}}\right)^{2}}}\\
=\ln(r).\label{eq:derivative_of_f}
\end{multline}
We now investigate this equation for $\frac{tv_{h}}{a}\ll1$.

First, we re-write this expression as a self-consistent equation 
\begin{equation}
m_{\max}=\frac{tv_{h}}{a}\sin\left(ak_{0}+\Delta(m_{\max})\right),
\end{equation}
where $\Delta$ is a function of $m$ defined by
\begin{equation}
\Delta(m_{\max})\equiv\frac{1}{2(\sigma/a)^{2}}\frac{tv_{h}}{a}\ln(r)\sqrt{1-\left(\frac{a}{tv_{h}}m_{\max}\right)^{2}}.
\end{equation}
We can solve this equation iteratively, order by order, in $\frac{tv_{h}}{a}$.
The first order is just given by
\begin{equation}
m_{1}=\sin(ak_{0})\frac{tv_{h}}{a}.
\end{equation}
Recalling that $\sin(ak_{0}+\Delta)\approx\sin(ak_{0})+\Delta\cos(ak_{0})-\frac{1}{2}\Delta^{2}\sin(ak_{0})$,
we see that the second order is given by 
\begin{equation}
m_{2}=\frac{1}{2(\sigma/a)^{2}}\cos^{2}(ak_{0})\ln(r)\left(\frac{tv_{h}}{a}\right)^{2}.
\end{equation}
The third order ($\propto\left(\frac{tv_{h}}{a}\right)^{3}$) will
be obtained as the sum of two terms:
\begin{equation}
m_{3}=m_{3,A}+m_{3,B},
\end{equation}
where the first term $m_{3,A}$ is $\left(\frac{tv_{h}}{a}\right)\Delta_{2}\cos(ak_{0})$,
where $\Delta_{2}\propto\left(\frac{tv_{h}}{a}\right)^{2}$ is the
second order term in the expansion of $\Delta$ in powers of $\frac{tv_{h}}{a}$,
and the second term $m_{3,B}$ comes from the second order of the
Taylor expansion, $-\left(\frac{tv_{h}}{a}\right)\frac{1}{2}\Delta_{1}^{2}\sin(ak_{0})$,
where $\Delta_{1}\propto\left(\frac{tv_{h}}{a}\right)$ is the first
order. A short calculation leads to
\begin{multline*}
m_{3,A}=2m_{3,B}=\\
-\frac{1}{4(\sigma/a)^{4}}\cos^{2}(ak_{0})\sin(ak_{0})\ln^{2}(r)\left(\frac{tv_{h}}{a}\right)^{3},
\end{multline*}
and therefore we have
\begin{equation}
m_{3}=-\frac{3}{8(\sigma/a)^{4}}\cos^{2}(ak_{0})\sin(ak_{0})\ln^{2}(r)\left(\frac{tv_{h}}{a}\right)^{3}.
\end{equation}

\section{\label{subsec:appendix_The-continuum-limit}The continuum limit of
the Hatano-Nelson model}

We start with the Hamiltonian
\begin{equation}
H_{0}=\sum_{n=1}^{N_{0}}t_{l,0}\left|na_{0}\right\rangle \left\langle na_{0}+a_{0}\right|+t_{r,0}\left|na_{0}+a_{0}\right\rangle \left\langle na_{0}\right|,
\end{equation}
where $a_{0}$ is the lattice spacing (initially, before taking the
continuum limit), and the total system size is $L=N_{0}a_{0}$. We
recall that the transformation (\ref{eq:Transformation}) will make
the Hamiltonian Hermitian, where $r_{0}=\left(\frac{t_{r,0}}{t_{l,0}}\right)^{1/2}$.
We will now consider the continuum limit by taking $x=na$, where
$a\rightarrow0$ (and $n\rightarrow\infty$ for any finite $x$) .
Since we expect that after the transformation $r^{n}\psi(na)$ will
converge (for $a\rightarrow0$) to a finite limit only as a function
of $x=na$, then we can write $r_{0}^{n_{0}}=r^{n}$ where $n_{0}a_{0}=na$,
and therefore
\begin{equation}
r=\left(\frac{t_{r,0}}{t_{l,0}}\right)^{a/2a_{0}}.
\end{equation}
We now take
\begin{equation}
t_{l}=\frac{1}{r}\sqrt{t_{l}t_{r}}\approx\sqrt{t_{l}t_{r}}\left(1-\frac{a}{2a_{0}}\ln\left(\frac{t_{r,0}}{t_{l,0}}\right)\right),
\end{equation}
\begin{equation}
t_{r}=r\sqrt{t_{r}t_{l}}\approx\sqrt{t_{l}t_{r}}\left(1+\frac{a}{2a_{0}}\ln\left(\frac{t_{r,0}}{t_{l,0}}\right)\right),
\end{equation}
where the approximation is valid for $a\rightarrow0$. Now, we define
as usual $\left|x\right\rangle \equiv\frac{1}{\sqrt{a}}\left|na\right\rangle $
and transform the sum into integral, resulting in
\begin{equation}
H=\int_{0}^{L}\left[t_{l}\left|x\right\rangle \left\langle x+a\right|+t_{r}\left|x+a\right\rangle \left\langle x\right|\right]dx.
\end{equation}
Approximating $\left|x+a\right\rangle \approx\left|x\right\rangle +a\partial_{x}\left|x\right\rangle +\frac{a^{2}}{2}\partial_{x}^{2}\left|x\right\rangle $
and assuming periodic boundary conditions, we get that
\begin{equation}
H=H_{\mathrm{h}}+H_{\mathrm{nh}},
\end{equation}
where $H_{\mathrm{h}},H_{\mathrm{nh}}$, the Hermitian and non-Hermitian
parts, are given by
\begin{equation}
H_{\mathrm{h}}=\sqrt{t_{l}t_{r}}\int_{0}^{L}\left(2\left|x\right\rangle \left\langle x\right|-a^{2}\left|x\right\rangle \partial_{x}^{2}\left\langle x\right|\right),
\end{equation}
\begin{equation}
H_{\mathrm{nh}}=\frac{a}{a_{0}}\ln\left(\frac{t_{r,0}}{t_{l,0}}\right)\sqrt{t_{l}t_{r}}\int_{0}^{L}\left|x\right\rangle \partial_{x}\left\langle x\right|.
\end{equation}
Therefore, the continuum Hamiltonian is given by
\begin{equation}
H=E_{0}-\frac{1}{2m}\partial_{x}^{2}+b\partial_{x},\label{eq:Hamiltonian_of_continuum_limit}
\end{equation}
where $E_{0}=2\sqrt{t_{l}t_{r}}$, $m=1/\left(2a^{2}\sqrt{t_{l}t_{r}}\right)$,
and $b=\frac{a^{2}}{a_{0}}\ln\left(\frac{t_{r,0}}{t_{l,0}}\right)\sqrt{t_{l}t_{r}}$.
Since $m,b$ should be finite, while taking the limit $a\rightarrow0$
we need to keep 
\begin{equation}
\sqrt{t_{l}t_{r}}=\frac{a_{0}^{2}}{a^{2}}\sqrt{t_{l,0}t_{r,0}},
\end{equation}
and therefore we finally get
\begin{equation}
m=\frac{1}{2a_{0}^{2}\sqrt{t_{l,0}t_{r,0}}},\,b=a_{0}\ln\left(\frac{t_{r,0}}{t_{l,0}}\right)\sqrt{t_{l,0}t_{r,0}}.\label{eq:appendix_expressions_for_m_b}
\end{equation}

\section{\label{subsec:appendix_non_hermitian_transition_time}Calculation
of the transition time of the non-Hermitian reflection}

In Section \ref{subsec:Non-Hermitian-Reflection} we showed that in
the presence of a wall, the wave-packet could experience a sadden
change in its behavior, related to the reflection event of the Hermitian
wave-packet. Here, we will show that in the continuum limit, the time
of this event (i.e., the time when the non-Hermitian wave-packet changes
its velocity), is completely determined by $t_{\mathrm{hit}}$, the
time that it takes to the \textit{Hermitian} wave-packet to hit the
wall. On the lattice, however, the time is larger than $t_{\mathrm{hit}}$.
We will also calculate here the first correction to this time in orders
of $a$, the lattice spacing.

We will start by using our approximation (Eq. \ref{eq:eq:appendix_final_approximation_simpler}),
to write the wave-packet as
\begin{equation}
\left|\left\langle m|\psi(t)\right\rangle \right|^{2}\approx e^{-2\frac{\sigma^{2}}{a^{2}}\left(\arcsin\left(\frac{am}{tv_{h}}\right)-ak_{0}\right)^{2}+2m\ln(r)},
\end{equation}
where we assume that the center at $t=0$ is $m=0$. We are interested
in the amplitude of this expression at the peak point, $m_{\max}$.
Using Eq. (\ref{eq:derivative_of_f}), we see that
\begin{multline*}
2(\sigma/a)^{2}\left(\arcsin\left(\frac{a}{tv_{h}}m_{\max}\right)-ak_{0}\right)=\\
\frac{\sqrt{1-\left(\frac{am_{\max}}{tv_{h}}\right)^{2}}}{\frac{a}{tv_{h}}}\ln(r),
\end{multline*}
and therefore we get
\begin{equation}
\left|\left\langle m_{\max}|\psi(t)\right\rangle \right|^{2}\approx e^{-\frac{\sqrt{1-\left(\frac{am_{\max}}{tv_{h}}\right)^{2}}}{\frac{a}{tv_{h}}}\ln(r)+2m_{\max}\ln(r)}.
\end{equation}
Since we assumed that the lattice spacing is small, the transition
must occur at small times as well. Thus, we write $m_{\max}(t)\approx\sin(ak_{0})\frac{tv_{h}}{a}$
and therefore $\frac{\sqrt{1-\left(\frac{am_{\max}}{tv_{h}}\right)^{2}}}{\frac{a}{tv_{h}}}=2t\cos(ak_{0})\sqrt{t_{r}t_{l}}\approx0$,
so we get that $\left|\left\langle m_{\max}|\psi(t)\right\rangle \right|^{2}\approx e^{2m_{\max}(t)\ln(r)}$.
Since the amplitude of the peak is given by the exponent of $m_{\max}$,
the time at which the amplitude of the image wave-packet surpasses
that of the incident wave-packet (and therefore becomes more dominant)
is the time when the maximum position $m_{\max}$ of the former surpasses
that of the latter. Therefore, we will write $m_{\max}$ for both
wave-packets and set them equal.

We can expand $m_{\max}$ for small times as
\begin{equation}
\Delta m_{\max}(t,k_{0})\approx A(k_{0})t+B(k_{0})t^{2}+C(k_{0})t^{3},
\end{equation}
where $\Delta m_{\max}(t,k_{0})=m_{\max}(t,k_{0})-m_{0}$, and the
coefficients are, respectively (see Appendix \ref{appendix_sbusec_small_and_large_t}),
\begin{flalign*}
A(k_{0}) & =\sin(ak_{0})\frac{v_{h}}{a},\\
B(k_{0}) & =\frac{\ln(r)}{2(\sigma/a)^{2}}\cos^{2}(ak_{0})\left(\frac{v_{h}}{a}\right)^{2},\\
C(k_{0}) & =-\frac{3\ln^{2}(r)}{8(\sigma/a)^{4}}\cos^{2}(ak_{0})\sin(ak_{0})\left(\frac{v_{h}}{a}\right)^{3}.
\end{flalign*} We note that
$A$,$C$ are odd in $k_{0}$, while $B$ is even. Assuming that the
initial distance from the wall is $d$, the maximum position of the
incident and the image wave-packet are respectively given by
\begin{equation}
m_{\max}^{\mathrm{inc}}(t)=-\frac{d}{a}+\Delta m_{\max}(t,k_{0}),
\end{equation}
\begin{equation}
m_{\max}^{\mathrm{img}}(t)=\frac{d}{a}+\Delta m_{\max}(t,-k_{0}).
\end{equation}
We wish to find the transition time $t$ where these expressions become
equal. Comparing them we find
\begin{equation}
\Delta m_{\max}(t,k_{0})-\Delta m_{\max}(t,-k_{0})=2\frac{d}{a},
\end{equation}
which leads to
\begin{equation}
A(k_{0})t+C(k_{0})t^{3}=\frac{d}{a}.\label{eq:transition_time_equation}
\end{equation}
For small lattice spacing $C(k_{0})$ will be small in comparison
to $A(k_{0})$, and we get to first order that
\begin{equation}
t_{\mathrm{hit}}=\frac{d/a}{A(k_{0})}=\frac{d}{\sin(ak_{0})v_{h}}.\label{eq:t_hit_A}
\end{equation}
This is exactly the time it takes to the Hermitian wave-packet to
hit the wall, as expected. Plugging $t=t_{\mathrm{hit}}+t_{\delta}$
into Eq. (\ref{eq:transition_time_equation}) where $t_{\delta}$
is the difference from the Hermitian transition time, we find that
the leading order in $a$ is
\begin{equation}
t_{\delta}\approx\frac{t_{\mathrm{hit}}^{3}}{-A(k_{0})/C(k_{0})}.\label{eq:t_hit_B}
\end{equation}
Keeping $a_{0}$ to be some reference lattice constant (where $a\ll a_{0}$),
we have $v_{h}=2a\sqrt{t_{r}t_{l}}$ where $\sqrt{t_{l}t_{r}}=\frac{a_{0}^{2}}{a^{2}}\sqrt{t_{l,0}t_{r,0}},$
and $r=\left(\frac{t_{r,0}}{t_{l,0}}\right)^{a/2a_{0}}$ (see Appendix
\ref{subsec:appendix_The-continuum-limit}). Plugging this into Eq.
(\ref{eq:t_hit_A},\ref{eq:t_hit_B}), we finally obtain
\begin{equation}
t_{\mathrm{hit}}=\frac{d}{\frac{\sin(ak_{0})}{a}a_{0}^{2}\sqrt{t_{l,0}t_{r,0}}},
\end{equation}
\begin{equation}
t_{\delta}\approx\frac{3}{64}\frac{\ln^{2}\left(\frac{t_{r,0}}{t_{l,0}}\right)}{\sqrt{t_{l,0}t_{r,0}}}\frac{\cos^{2}(ak_{0})}{\frac{\sin(ak_{0})^{3}}{a^{3}}}\frac{d^{3}a^{2}}{a_{0}^{4}\sigma^{4}}.
\end{equation}

\subsection{\label{subsec:The-effects-of_sigma_d_on_transition}The effects of
$\sigma$ and $d$ on the transition}

\begin{figure*}
\centerline{\includegraphics[scale=0.7]{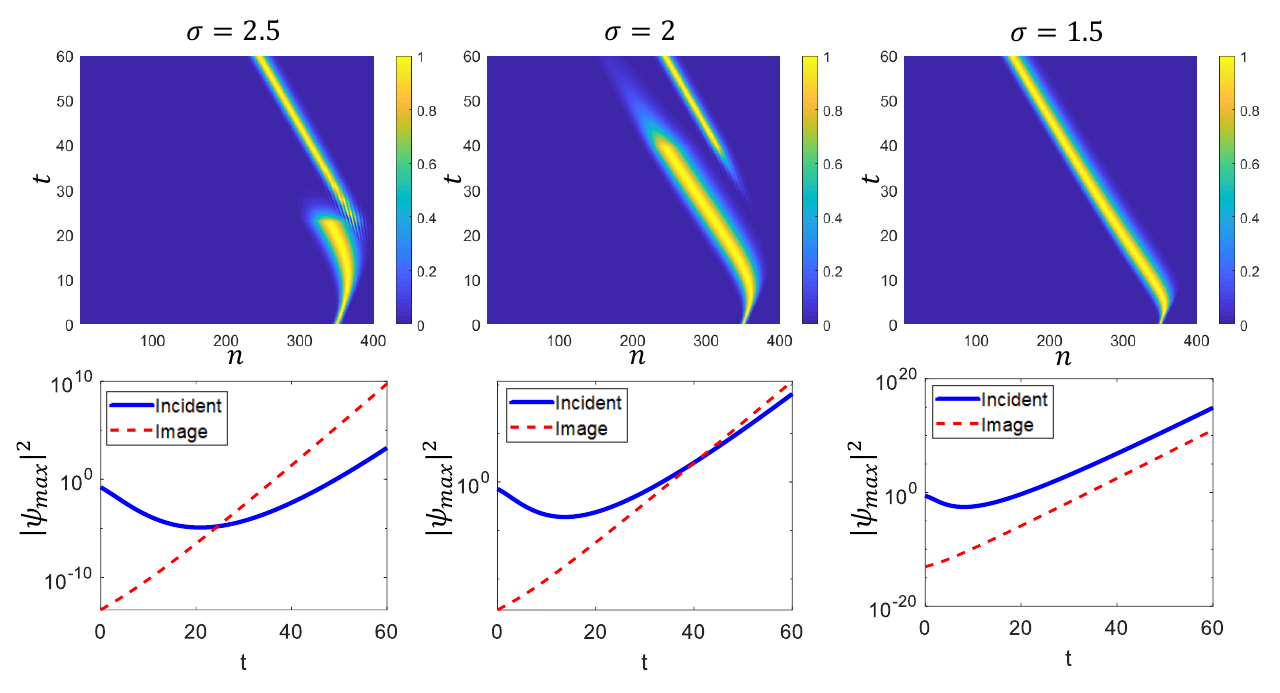}}

\caption{\label{fig:appendix4_A}Top panels: the dynamics of the non-Hermitian
wavepacket, with parameters identical to those of Fig. \ref{fig:gaussian_dynamics_reflection},
but with initial distance from the wall $d=50$ (instead of 100),
and for three different values of the initial widths $\sigma=2.5,2,1.5$.
Bottom panels: The amplitude of the peak of the Gaussian, for the
incident wave-packet (in blue) and its image (red dotted-line). We
can see that $1.5<\sigma_{c,\mathrm{ref}}<2$.}
\end{figure*}

\begin{figure*}
\centerline{\includegraphics[scale=0.7]{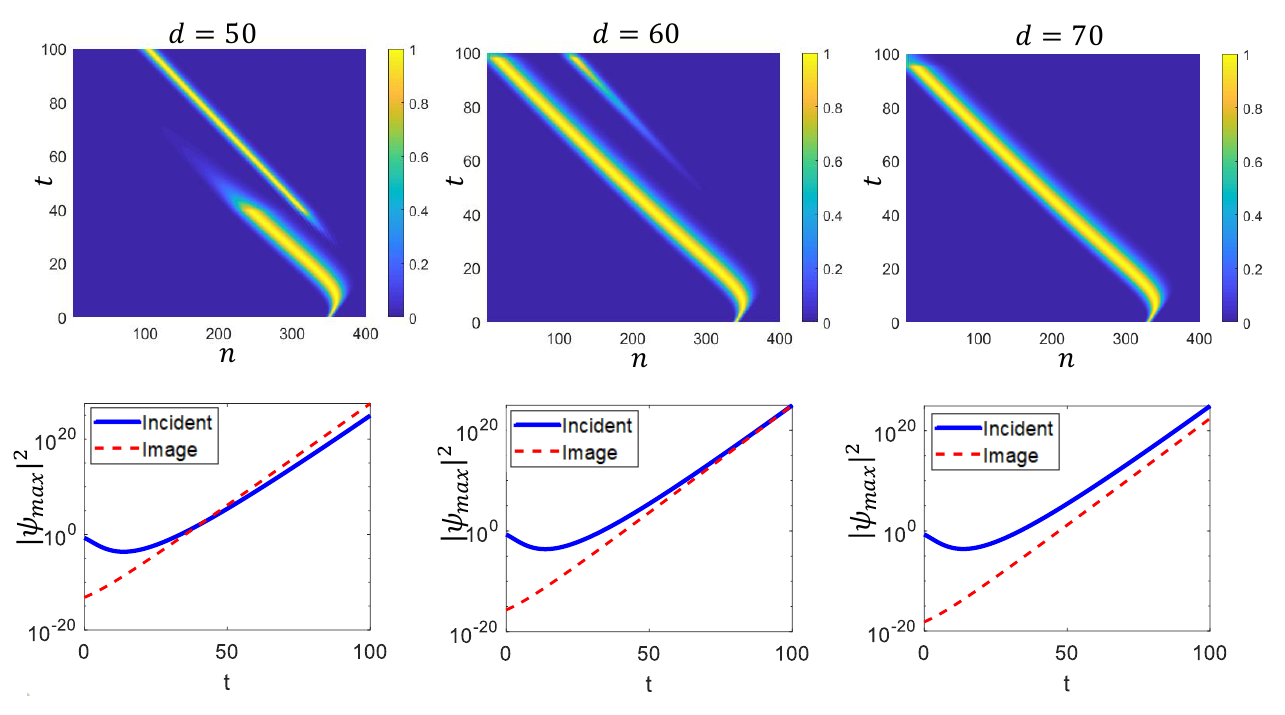}}

\caption{\label{fig:appendix4_B}Top panels: the dynamics of the non-Hermitian
wavepacket, with parameters identical to those of Fig. \ref{fig:gaussian_dynamics_reflection},
but with $\sigma=2$, and for three
different values of initial distance from the wall, $d=50,60,70$.
Bottom panels: The amplitude of the peak of the Gaussian, for the
incident wave-packet (in blue) and its image (red dotted-line).}
\end{figure*}

As described in the main text, the width $\sigma$ and
the initial distance from the wall $d$ have a significant effect
on whether the transition will occur (i.e., whether the image wave-packet
becomes more dominant than the original wave-packet). In Fig. \ref{fig:appendix4_A},
we present the dynamics for three values of $\sigma$. We observe
that $\sigma=2.5$ and $\sigma=2$ are above the critical threshold
(a transition is visible), while $\sigma=1.5$ is below it. The reason
for this can be identified in the bottom panels of the figure, where
we plot the amplitude of the peaks of the original and image wave-packets.
The intersection point represents the time when the image wave-packet
becomes more dominant than the original wave, and thus, a transition
is observed. We can see that reducing $\sigma$ causes the acceleration
toward the limiting velocity (which corresponds to the growth rate)
to increase, until at some point the intersection no longer occurs.
In Fig. \ref{fig:appendix4_B} , we observe a similar effect when
$\sigma$ is held constant, but the distance $d$ from the wall is
varied. We see that the critical value of $\sigma$ depends on $d$.

To determine this critical value, we can use the approximation derived
above. The peak position can be found by solving Eq. (\ref{eq:derivative_of_f}),
and the amplitude can be obtained by substituting this result into
Eq. (\ref{eq:approximation_appendix}). We can then calculate the
growth of both the real and image waves, noting that the initial amplitude
of the image wave-packet is lower by a factor of $r^{2d}$, and that its
initial momentum is $-k_{0}$ rather than $k_{0}$. Finally, we can
check the minimum value of $\sigma$ for which such an intersection
occurs.

\section{\label{subsec:Appendix_Disorder}Effect of disorder: Supplemental
details}

\subsection{\label{sec:Localization-length-estimation}Localization length estimation}

In Section \ref{subsec:Disorder} we discussed the effect of disorder,
by considering an onsite disorder term $\sum_{n}w_{n}\left|n\right\rangle \left\langle n\right|$,
where $w_{n}$ are uniformly distributed in $[-w,w]$. Here, we will
provide a simple estimation for the localization length as a function
of $w$ and show that $\xi\propto W^{-2}$. Therefore, for the value
that were chosen in Section \ref{subsec:Disorder} ($L=400a,$ $w=10^{-7}$),
we get that $\xi\gg L$ indicating that the disorder is weak and does
not lead to localization. Starting from the Fermi's golden rule for
a time-independent potential, we have
\begin{equation}
\frac{1}{\tau_{i\rightarrow f}}=2\pi|\left\langle f|V|i\right\rangle |^{2}\rho(E_{f})\delta(E_{f}-E_{i}),
\end{equation}
where $\tau_{i\rightarrow f}$ is the scattering time between states
$i,f$. Taking $\left|i,f\right\rangle =\frac{1}{\sqrt{N}}\sum_{n}e^{ik_{i,f}n}\left|n\right\rangle $
and $\left\langle n'|V|n\right\rangle =V(n)\delta_{n,n'}$, we get
\begin{equation}
\left|\left\langle f|V|i\right\rangle \right|^{2}=\frac{1}{N^{2}}\sum_{n,n'}e^{i\left(k_{i}-k_{f}\right)\left(n-n'\right)}V(n)V(n').
\end{equation}
Denoting the average over disorder realizations by $\left\llbracket \right\rrbracket $, we get that $\left\llbracket V(n)V(n')^{*}\right\rrbracket =\frac{w^{2}}{3}\delta_{n,n'}$,
since $V(n)$ which is independent and uniformly distributed over
the interval $[-w,w]$. Therefore, we get $\left\llbracket \left|\left\langle f|V|i\right\rangle \right|^{2}\right\rrbracket =\frac{1}{N}\frac{w^{2}}{3}$,
so the total scattering time is $\frac{1}{\tau}=\int dE_{f}\frac{1}{\tau_{i\rightarrow f}}=2\pi\frac{1}{N}\frac{w^{2}}{3}\rho(E)$.
Plugging the density of states on the 1D lattice $\rho(E)=\frac{N}{2\pi t_{0}}\frac{1}{\sqrt{1-\frac{E^{2}}{4t_{0}^{2}}}},$
we find that 
\begin{equation}
\frac{1}{\tau}=\frac{w^{2}}{3}\frac{1}{t_{0}}\frac{1}{\sqrt{1-\frac{E^{2}}{4t_{0}^{2}}}}.
\end{equation}
We also note that the velocity on the lattice is $v=\frac{\partial E}{\partial k}=2at_{0}\sqrt{1-\frac{E^{2}}{4t_{0}^{2}}}.$
Finally, since in 1D the localization length $\xi$ is just the mean free
path, we get that
\begin{equation}
\xi(E)=\tau v=6a\frac{t_{0}^{2}}{w^{2}}\left(1-\frac{E^{2}}{4t_{0}^{2}}\right).
\end{equation}

\subsection{Analysis of disorder effects and calculation of the transition time}

In this section we will analyze the dynamics of a Gaussian wave-packet
with disorder. We will also estimate the transition time, which is
the time the wave-packet's peak change its position due to disorder,
as presented in Fig. \ref{fig:disorder_transition}. Again, we consider
the Hamiltonian $H=H_{0}+H',$where $H_{0}$ is given by Eq. (\ref{eq:Hanato_nelson}),
and the disorder term is given by $H'=\sum_{n}w_{n}\left|n\right\rangle \left\langle n\right|$,
where $w_{n}$ are uniformly distributed in $[-w,w]$. Any wave-function
can be written as
\begin{equation}
\left|\psi(t)\right\rangle =\sum_{k}c_{k}(t)e^{-iE_{k}t}\left|k\right\rangle ,
\end{equation}
where $\left|k\right\rangle $ is the eigenbasis of $H_{0}$ ($H_{0}\left|k\right\rangle =E_{k}\left|k\right\rangle $),
$c_{k}(t=0)=\left\langle \left.k\right|\psi(t=0)\right\rangle ,$
and $e^{-iE_{k}t}$ is the dynamic phase factor. Thus, we see that
\begin{multline*}
\frac{\partial c_{k}(t)}{\partial t}+\frac{i}{\hbar}\left\langle k\right|H'\left|k\right\rangle c_{k}(t)=\\
-\frac{i}{\hbar}\sum_{k'\neq k}c_{k'}(t)e^{-it\left(E_{k}-E_{k'}\right)/\hbar}\left\langle k\right|H'\left|k'\right\rangle .
\end{multline*}
Multiplying by the integration factor $e^{\frac{i}{\hbar}\left\langle k\right|H\left|k\right\rangle t}$
and taking the integral, we get
\begin{multline*}
e^{\frac{i}{\hbar}\left\langle k\right|H'\left|k\right\rangle t}c_{k}(t)=\\
c_{k}(0)-\int_{0}^{t}\frac{i}{\hbar}\sum_{k'\neq k}c_{k'}(t=0)e^{-\frac{i}{\hbar}t\left(E_{k}-E_{k'}-\left\langle k\right|H'\left|k\right\rangle \right)}\left\langle k\right|H'\left|k'\right\rangle .
\end{multline*}
To obtain the first-order behavior for small $t$, we assume that
$c_{k}(t)=c_{k}(t=0)$ in the sum and find
\begin{multline*}
c_{k}(t)=e^{-\frac{i}{\hbar}\left\langle k\right|H'\left|k\right\rangle t}c_{k}(0)+e^{-\frac{i}{\hbar}\left\langle k\right|H'\left|k\right\rangle t}\times\\
\times\sum_{k'}c_{k}(t=0)\frac{e^{-it\left(E_{k}-E_{k'}-\left\langle k\right|H'\left|k\right\rangle \right)/\hbar}-1}{E_{k}-E_{k'}-\left\langle k\right|H'\left|k\right\rangle }\left\langle k\right|H'\left|k'\right\rangle .
\end{multline*}

We would now want to take the disorder average of $|c_{k}(t)|^{2}.$
First we recall that $\left|k\right\rangle =\frac{1}{\sqrt{N}}\sum_{n}e^{-ikna}\left|n\right\rangle ,$
and therefore
\begin{multline*}
\left\langle k''\right|H'\left|k\right\rangle \left\langle k\right|H'\left|k'\right\rangle =\\
\frac{1}{N^{2}}\sum_{n,n'}w_{n}w_{n'}e^{i\left(k''-k\right)n'a+i\left(k-k'\right)na}.
\end{multline*}
Using the fact that $\left\llbracket w_{n}w_{n'}\right\rrbracket =\frac{w^{2}}{3}\delta_{n,n'}$ (disorder average),
we find
\begin{equation}
\left\llbracket \left\langle k''\right|H'\left|k\right\rangle \left\langle k\right|H'\left|k'\right\rangle \right\rrbracket =\frac{w^{2}}{3}\frac{1}{N}\delta_{k'',k'}.
\end{equation}
Using this, we obtain
\begin{multline}
\left\llbracket |c_{k}(t)|^{2}\right\rrbracket -|c_{k}(0)|^{2}=\\
\frac{w^{2}}{3}\frac{1}{N}\sum_{k'\neq k}|c_{k'}(t=0)|^{2}\left|\frac{e^{-it\left(E_{k}-E_{k'}\right)/\hbar}-1}{E_{k}-E_{k'}}\right|^{2},\label{eq:average_of_c_k}
\end{multline}
where we note that the cross-terms vanished since $\left\llbracket \left\langle k''\right|H\left|k\right\rangle \right\rrbracket =0$.
For small $t$ we have
\begin{equation}
\left\llbracket |c_{k}(t)|^{2}\right\rrbracket \approx|c_{k}(0)|^{2}+\frac{t^{2}}{\hbar^{2}}\frac{w^{2}}{3}\frac{1}{N}.
\end{equation}

Since the spectrum contains an imaginary term, for large $t$ we will
see an exponential growth. However, between the polynomial behavior
observed above for small times and the exponential growth regime,
there may be an intermediate time regime where $\left\llbracket |c_{k}(t)|^{2}\right\rrbracket $
stabilize at a certain value. We can estimate the saturation value
by neglecting $\mathrm{\mathrm{Im}}\left[E_{k}-E_{k'}\right]$, so
that Eq. (\ref{eq:average_of_c_k}) becomes
\begin{multline}
\left\llbracket |c_{k}(t)|^{2}\right\rrbracket -|c_{k}(0)|^{2}\approx\\
4\frac{w^{2}}{3}\frac{1}{N}\sum_{k'}|c_{k'}(t=0)|^{2}\left[\frac{\sin\left(\frac{t}{2\hbar}\mathrm{Re}(E_{k}-E_{k'})\right)}{\mathrm{Re}\left(E_{k}-E_{k'}\right)}\right]^{2},\label{eq:appendix_disorder_prediction_small_t}
\end{multline}
where we note that the approximation is true up to intermediate values
of $t$. The saturation will occur when the sine function becomes
of order unity, so we get that the saturation value is
\begin{equation}
\left\llbracket |c_{k}(t=t_{s})|^{2}\right\rrbracket \approx|c_{k}(0)|^{2}+4\frac{w^{2}}{3}\frac{1}{N}\frac{1}{\mathrm{Re}\left(E_{k}-E_{k_{0}}\right)^{2}},
\end{equation}
and the saturation time is
\begin{equation}
t_{s}\approx\frac{\pi}{\mathrm{Re}(E_{k}-E_{k_{0}})},
\end{equation}
where we also assumed that $c_{k}(t=0)\propto e^{-\sigma^{2}(k-k_{0})^{2}}$,
so that only $k'\approx k_{0}$ will contributed to the sum. Plugging
in $k_{0}=\frac{\pi}{4}$ and $k=-\pi/2$, we get that
\begin{equation}
\left\llbracket |c_{k=-\pi/2}(t=t_{s})|^{2}\right\rrbracket \approx\frac{8}{3}\frac{w^{2}}{N}\frac{1}{(t_{r}+t_{l})^{2}},\,t_{s}\approx\frac{\pi\sqrt{2}}{(t_{r}+t_{l})^{2}}.
\end{equation}
Now, to get an estimation for large times, we can finally use the
expression
\begin{equation}
\left\llbracket |c_{k=-\pi/2}(t)|^{2}\right\rrbracket \approx\frac{8}{3}\frac{w^{2}}{N}\frac{1}{(t_{r}+t_{l})^{2}}e^{2(t_{l}-t_{r})(t-t_{s})}.\label{eq:appendix_disorder_prediction_largel_t}
\end{equation}
\begin{figure}
\centerline{\includegraphics[scale=0.38]{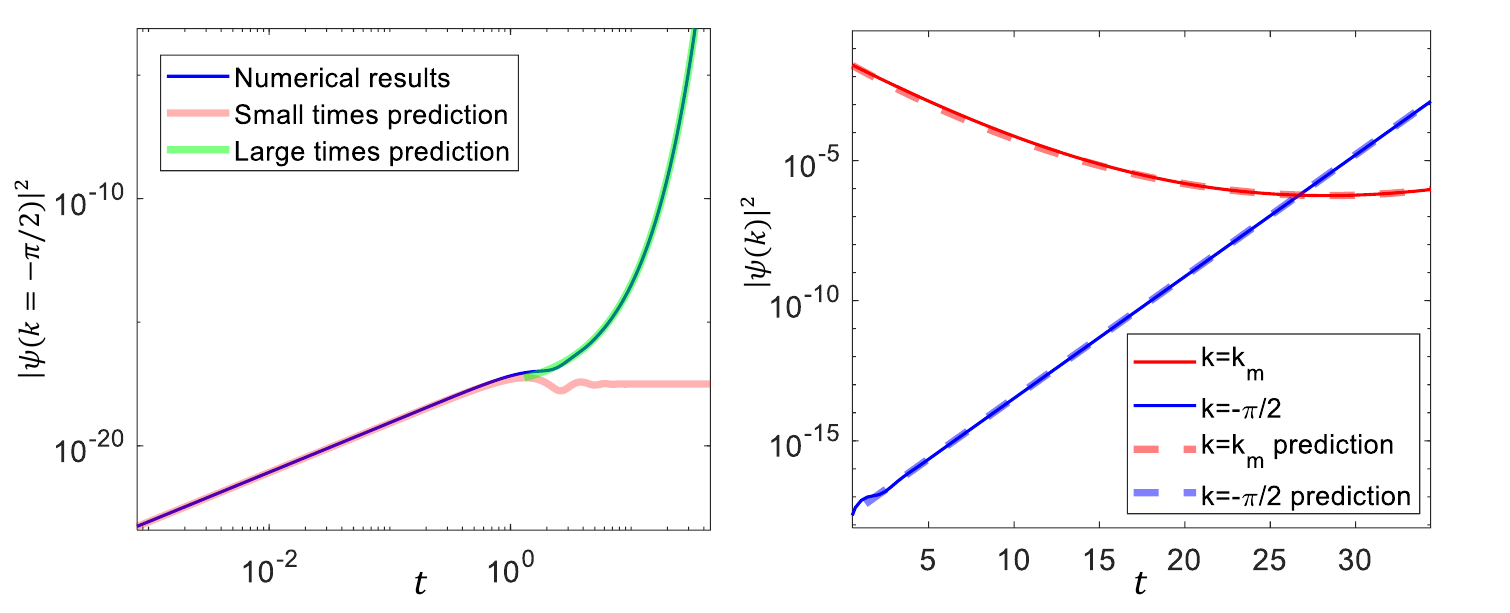}}

\caption{\label{fig:appendix_disorder_predictions}Left panel: The numerical
results versus the approximations for small and large times. Right
panel: The numerical results of the maximum of the wave-packet (in
red), and the numerical results for $k=-\pi/2$, versus their approximations
(in dotted lines). The setup is the same as in Fig. \ref{fig:disorder_transition}.}
\end{figure}
In the left panel of Fig. \ref{fig:appendix_disorder_predictions}
we compare Eq. (\ref{eq:appendix_disorder_prediction_small_t}),(\ref{eq:appendix_disorder_prediction_largel_t})
with the numerical results, and see that they provide a very good approximation
to the behavior of $\left\llbracket |c_{k=-\pi/2}(t)|^{2}\right\rrbracket $
for small and large times, respectively.

We can now find the transition time between the Gaussian wave-packet
to the wave-packet corresponding to the maximum growth ($k=-\pi/2$).
In order to do so, we need to find the time where the maximum magnitude
of the Gaussian wave-packet becomes equal to the $k=-\pi/2$ packet.
For $k=-\pi/2$, we will use our approximation from Eq. (\ref{eq:appendix_disorder_prediction_largel_t}).
As for the maximum of the Gaussian wave-packet, in order to simplify
the equations we will use the continuum limit given by Eq. (\ref{eq:Hamiltonian_of_continuum_limit}),
which, as we have seen, is a good approximation up to moderate times.
The spectrum is therefore
\begin{equation}
E=\frac{k^{2}}{2m}+ikb,
\end{equation}
where in our case (since $a=1$), $b,m$ are given by
\begin{equation}
m=\frac{1}{2\sqrt{t_{l}t_{r}}},\,b=\ln\left(\frac{t_{r}}{t_{l}}\right)\sqrt{t_{l}t_{r}}.
\end{equation}
Thus, we have
\begin{equation}
\psi(k,t)=\left(\frac{8\pi\sigma^{2}}{N^{2}}\right)^{1/4}e^{-\sigma^{2}(k-k_{0})^{2}}e^{-i\left(\frac{k^{2}}{2m}+ikb\right)t}.
\end{equation}
By completing the square, we get that
\begin{equation}
|\psi(k,t)|^{2}=\left(\frac{8\pi\sigma^{2}}{N^{2}}\right)^{1/2}e^{2k_{0}bt+\frac{b^{2}}{2\sigma^{2}}t^{2}}e^{-2\sigma^{2}\left[k-\left(k_{0}+\frac{1}{2}\frac{b}{\sigma^{2}}t\right)\right]^{2}},
\end{equation}
so that the maximum amplitude is given by
\begin{equation}
|\psi(k_{m},t)|^{2}=\left(\frac{8\pi\sigma^{2}}{N^{2}}\right)^{1/2}e^{2k_{0}bt+\frac{b^{2}}{2\sigma^{2}}t^{2}}.\label{eq:disorder_k_Gaussian_maximum_amplitude}
\end{equation}
That is, to determine the transition time due to disorder, we need
to find when Eq. (\ref{eq:appendix_disorder_prediction_largel_t})
equals Eq. (\ref{eq:disorder_k_Gaussian_maximum_amplitude}). This
yields a simple quadratic equation, and its solution gives the transition
time. In the right panel of Fig. (\ref{fig:appendix_disorder_predictions})
we present the numerical results for the amplitude of $k=k_{m}$ and
$k=-\pi/2$, and show that they are in a good agreement with the approximations
in Eq. (\ref{eq:appendix_disorder_prediction_largel_t}), (\ref{eq:disorder_k_Gaussian_maximum_amplitude}).

\subsubsection{The critical width $\sigma_{c,\mathrm{dis}}$ }

\begin{figure}
\centerline{\includegraphics[scale=0.45]{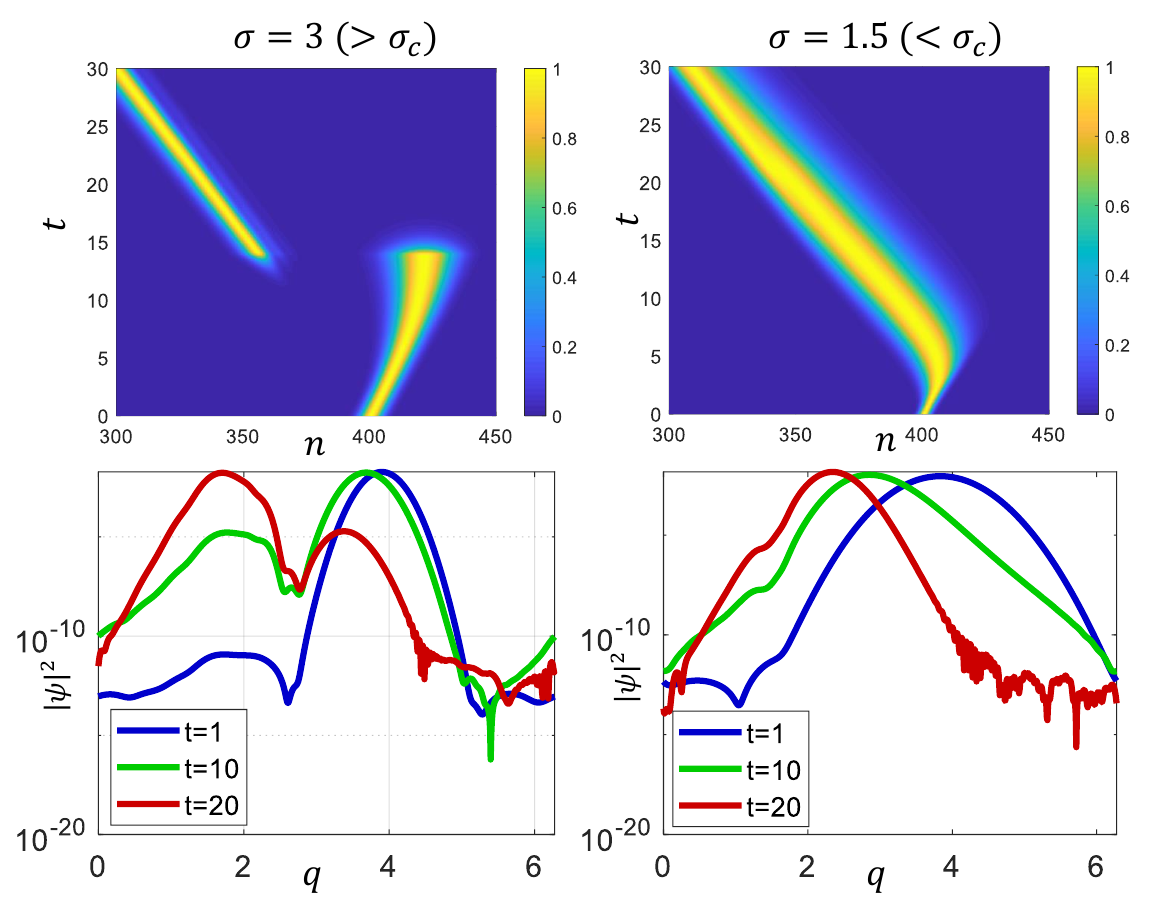}}

\caption{\label{fig:appendix_disorder_critical_value}Top panels: The transition
for $w=10^{-4}$ and two different values of $\sigma$ (below and
above the critical value). Bottom panels: The wave-pakcet in $q$-space
for at three different times, $t=1,10,20$. Except for $\sigma,w$,
the setup is the same as in Fig. \ref{fig:disorder_transition}.}
\end{figure}

\begin{figure}
\centerline{\includegraphics[scale=0.45]{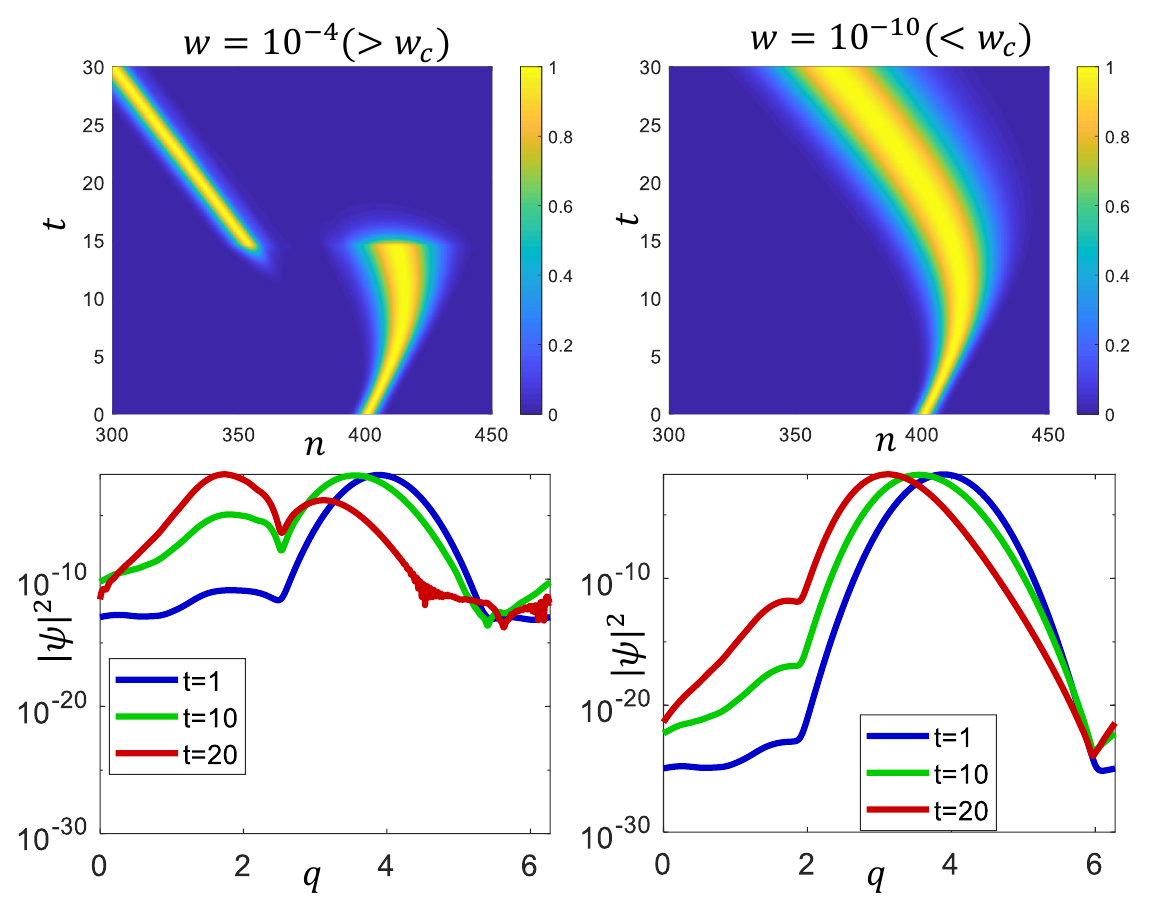}}

\caption{\label{fig:appendix_disorder_critical_value_B}Top panels: The transition
for $\sigma=2.5$ and two different values of $w$ (below and above
the critical value). Bottom panels: The wave-pakcet in $q$-space
for at three different times, $t=1,10,20$. Except for $\sigma,w$,
the setup is the same as in Fig. \ref{fig:disorder_transition}.}
\end{figure}

As discussed in the main text, the transition can only be observed
within a specific range of values for $\sigma,w$. Specifically, we
show in Fig. \ref{fig:appendix_disorder_critical_value} (\ref{fig:appendix_disorder_critical_value_B})
that for a fixed $w$ ($\sigma$), there is a critical value for $\sigma$
($w$). 
 
\section{\label{subsec:Appendix_beyond_HN}Additional results in the non-Hermitian
SSH model}

this section, we briefly discuss the dynamics of the non-Hermitian
SSH model (Eq. (\ref{eq:non-Hermitian_SSH})) using an initial Gaussian
wave-packet.
\begin{figure}
\centerline{\includegraphics[scale=0.45]{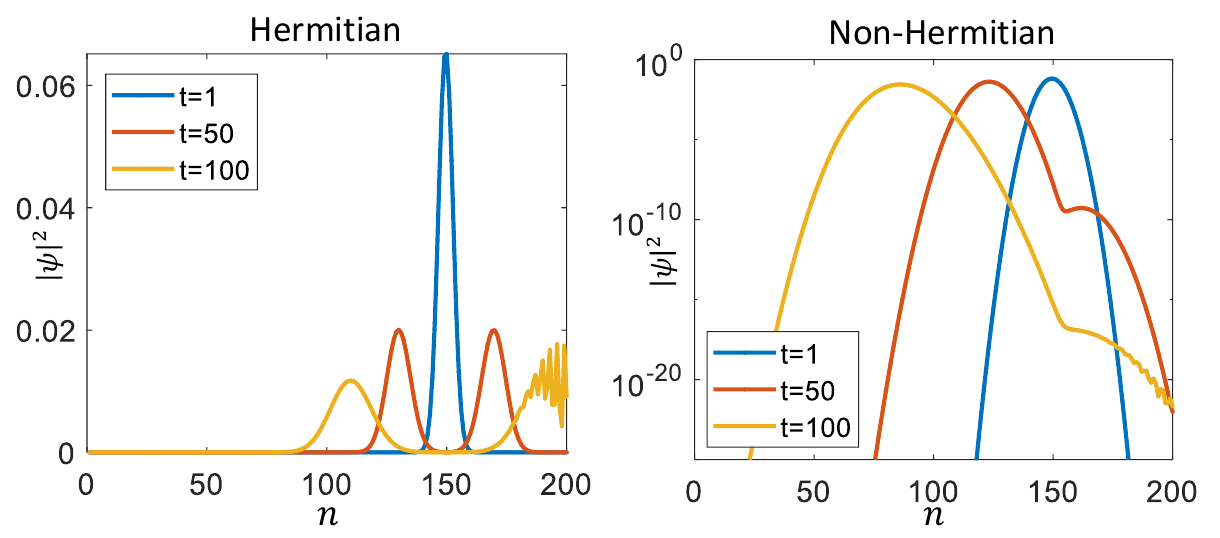}}

\caption{\label{fig:appendix_non_hermitian_SSH_dynamics}The dynamics of the
Hermitian (left panel) and non-Hermitian (right panel) wave-packets
(with and without the transformation in Eq. (\ref{eq:non_hermitian_SSH_transformation})),
in the non-Hermitian SSH model. The initial conditions are $\psi(n,A,t=0)=\frac{1}{\sqrt{4\pi\sigma^{2}}}e^{-\frac{a^{2}\left(n-n_{0}\right)^{2}}{4\sigma^{2}}+ik_{0}an}$
and $\psi(n,B,t=0)=0$, and the results are presented at 3 different
times ($t=1,50,100$). The parameters are $N=200$, $a=1$, $t_{1}=0.4,$
$t_{2}=1,$ $\gamma=0.5$, $n_{0}=150$, $\sigma=3,$ and $k_{0}=\pi/4$.}
\end{figure}
Fig. \ref{fig:appendix_non_hermitian_SSH_dynamics} illustrates the
wave-packet dynamics in the non-Hermitian SSH model, both with and
without the transformation given in Eq. (\ref{eq:non_hermitian_SSH_transformation}).
In the Hermitian case, the wave-packet splits into two Gaussians moving
in opposite directions. In the non-Hermitian model, we also observe
two Gaussian wave-packets, where each of them could be investigated
by similar tools that were used in Sec. \ref{sec:Non-Hermitian-wave-packet-dynami}.
That is, each one of them will accelerate to the left before saturating
at a finite velocity. However, due to the non-Hermiticity of the model,
the leftward-moving packet is amplified, making the rightward-moving
packet challenging to observe (at at least for $n<n_{0}$), as shown
in the figure. Specifically, the transition discussed in Sec. \ref{subsec:Non-Hermitian-Reflection}
(resulting from the reflection of the Hermitian wave) can also occur,
but will be hard to observe. In contrast, disorder-induced transitions
(like those investigated in Sec. \ref{subsec:Disorder}) can easily
be seen, as is shown in Fig. \ref{fig:appendix_non_hermitian_SSH_dynamics_disorder}.
\begin{figure}
\centerline{\includegraphics[scale=0.45]{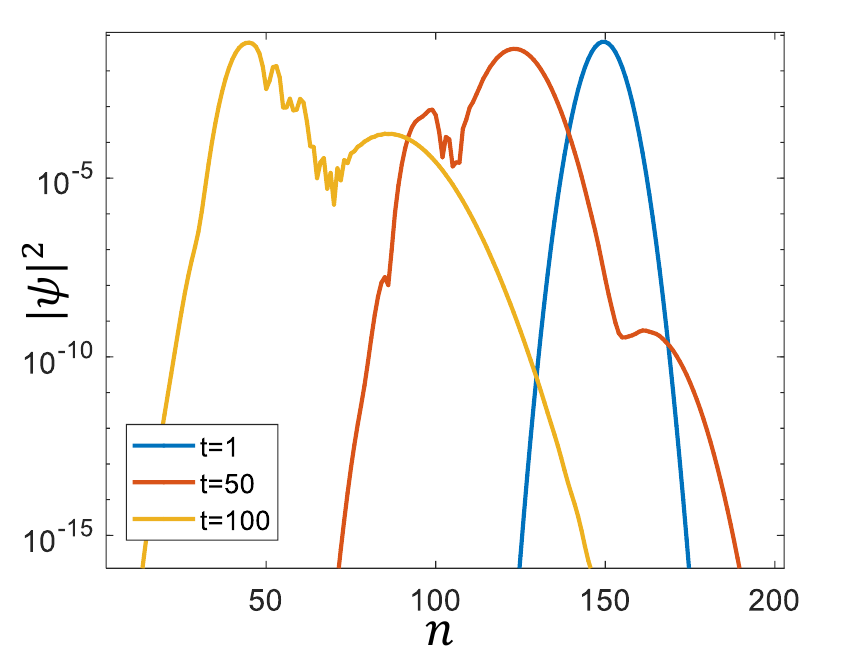}}

\caption{\label{fig:appendix_non_hermitian_SSH_dynamics_disorder}The dynamics
of the non-Hermitian SSH model with disorder. The setup is identical
to that of Fig. \ref{fig:appendix_non_hermitian_SSH_dynamics}, but
with additional onsite disorder term with $w=0.001$.}
\end{figure}
\bibliography{bibi}

\end{document}